\documentclass[11pt,a4paper]{article}
\pdfoutput=1

\usepackage{amsmath,amsfonts,mathrsfs,graphicx,bbm,bm}
\usepackage{fixltx2e}
\usepackage{cancel}

\usepackage[font=small]{caption}
\usepackage[text={450pt,650pt},headheight={14.5pt},centering]{geometry}

\usepackage{lmodern}

\usepackage{graphicx}
\usepackage{booktabs}
\usepackage{dsfont}

\usepackage{mathrsfs}
\usepackage{amsmath,amssymb}
\usepackage{mathtools}
\usepackage{bbm}
\usepackage{slashed}
\usepackage{braket}

\usepackage{comment}
\usepackage{hyperref}

\usepackage{xspace}
\hypersetup{
    colorlinks,
    linkcolor={red!0!black},
    citecolor={green!0!black},
    urlcolor={blue!0!black}
}

\usepackage{color}
\usepackage{tikz}

\definecolor{grey}{rgb}{0.4,0.4,0.5}
\definecolor{darkgreen}{rgb}{0,0.5,0}
\definecolor{darkred}{rgb}{0.6,0.0,0}
\definecolor{lightbrown}{rgb}{1,0.9,0.8}
\definecolor{brown}{rgb}{0.6,0.3,0.3}
\definecolor{darkblue}{rgb}{0,0,0.8}
\definecolor{darkmagenta}{rgb}{0.5,0,0.5}

\numberwithin{equation}{section}

\makeatletter
 \let\old@startsection=\@startsection
 \let\oldl@section=\l@section
 \renewcommand{\@startsection}[6]{\old@startsection{#1}{#2}{#3}{#4}{#5}{#6\mathversion{bold}}}
 \renewcommand{\l@section}[2]{\oldl@section{\mathversion{bold}#1}{#2}}
\makeatother

\renewcommand{\leq}{\leqslant}
\renewcommand{\geq}{\geqslant}

\def\XXint#1#2#3{{\setbox0=\hbox{$#1{#2#3}{\int}$}
    \vcenter{\hbox{$#2#3$}}\kern-.5\wd0}}

\newcommand{\alg}[1]{\mathfrak{#1}}

\def\be{\begin{equation}}
\def\ee{\end{equation}}

\newcommand{\bea}{\begin{equation}\begin{aligned}}
\newcommand{\eea}{\end{aligned}\end{equation}}
\newcommand{\bei}{\begin{itemize}}
\newcommand{\eei}{\end{itemize}}
\newcommand{\bee}{\begin{enumerate}}
\newcommand{\eee}{\end{enumerate}}
\newcommand{\bal}{\begin{equation}\begin{aligned}}
\newcommand{\eal}{\end{aligned}\end{equation}}

\def\a {\alpha}
\def\b {\beta}

\newcommand{\su}{\alg{su}}

\newcommand{\psu}{\alg{psu}}

\newcommand{\gen}{\mathbf}
\newcommand{\bpsi}{\bm{\psi}}

\renewcommand{\sin}{\text{sin}}
\renewcommand{\cos}{\text{cos}}

\newcommand{\dsbl}{[\![}
\newcommand{\dsbr}{]\!]}

\newcommand{\adjustedaccent}[1]{%
  \mathchoice{}{}
    {\mbox{\raisebox{-.5ex}[0pt][0pt]{$\scriptstyle#1$}}}
    {\mbox{\raisebox{-.35ex}[0pt][0pt]{$\scriptscriptstyle#1$}}}
}
\newcommand\barup[1]{\overset{\adjustedaccent{\smallsmile}}{#1}}
\newcommand\bardown[1]{\overset{\adjustedaccent{\smallfrown}}{#1}}

\begin{document}

\thispagestyle{empty}

\begin{flushright}\footnotesize\ttfamily
DMUS-MP-17/07\\
NORDITA 2017-066
\end{flushright}
\vspace{5em}

\begin{center}
\textbf{\Large\mathversion{bold} $q$-Poincar\'e supersymmetry in $AdS_5/CFT_4$}

\vspace{2em}

\textrm{\large Riccardo Borsato${}^1$ and Alessandro Torrielli${}^{2}$} 

\vspace{2em}

\begingroup\itshape
1. Nordita, Stockholm University and KTH Royal Institute of Technology,\\ Roslagstullsbacken 23, SE-106 91 Stockholm, Sweden\\[0.2cm]

2. Department of Mathematics, University of Surrey, Guildford, GU2 7XH, UK\par\endgroup

\vspace{1em}

\texttt{riccardo.borsato@su.se, a.torrielli@surrey.ac.uk}


\end{center}

\vspace{6em}

\begin{abstract}\noindent
We consider the exact S-matrix governing the planar spectral problem for strings on $AdS_5\times S^5$ and $\mathcal N=4$ super Yang-Mills, and we show that it is invariant under a novel ``boost'' symmetry, which acts as a differentiation with respect to the particle momentum. This generator leads us also to reinterpret the usual centrally extended $\mathfrak{psu}(2|2)$ symmetry, and to conclude that the S-matrix is invariant under a $q$-Poincar\'e supersymmetry algebra, where the deformation parameter is related to the 't Hooft coupling. We determine the two-particle action (coproduct) that turns out to be non-local, and study the property of the new symmetry under crossing transformations. We look at both the strong-coupling (large tension in the string theory) and weak-coupling (spin-chain description of the gauge theory) limits; in the former regime we calculate the cobracket utilising the universal classical $r$-matrix of Beisert and Spill. In the eventuality that the boost has higher partners, we also construct a quantum affine version of 2D Poincar\'e symmetry, by contraction of the quantum affine algebra $U_q(\widehat{\mathfrak{sl}_2})$ in Drinfeld's second realisation.
\end{abstract}

\newpage

\tableofcontents


\section{\label{sec:level1}Introduction}

\subsection{Exotic quantum groups in $AdS/CFT$}

Recent years have witnessed a dramatic progress in our understanding of the inner workings of the AdS/CFT correspondence, thanks to the discovery of integrability in planar ${\cal{N}}=4$ super Yang-Mills, and in the related  $AdS_5 \times S^5$ string sigma-model \cite{Beisert:2010jr,Arutyunov:2009ga}. The turning point of these developments consists in translating the associated spectral problem into the diagonalisation of the Hamiltonian of an effective two-dimensional integrable system, {traditionally solved by} Bethe-ansatz and exact S-matrix techniques. Symmetries play a crucial role in this program, and {their organisation into} the language of Hopf (super-)algebras {makes the problem amenable to a mathematical treatment in terms of algebraic relations and the corresponding representation theory}.
Such Hopf superalgebras turn out to be quite unconventional, and, as we will soon summarise, their properties are only partially understood. 

The symmetry {manifest} from the very formulation of the model is the superconformal algebra $\alg{psu}(2,2|4)$, {which is actually a {\it kinematical} symmetry sitting inside a much larger {\it dynamical} one. In fact,} it turns out that at the core of $AdS_5/CFT_4$ integrability there is an exotic Yangian-type algebra \cite{Drinfeld:1985rx,Drinfeld:1986in,Molev:1994rs,Khoroshkin:1994uk,Molev:2003,MacKay:2004tc,Beisert:2007ds,Torrielli:2011gg}, organised into levels labelled by a natural number. The level 0 coincides with the centrally-extended $\alg{psu}(2|2)$ Lie superalgebra first written down by Beisert \cite{Beisert:2005tm,Arutyunov:2006ak}. 
One of the first traits of unusual behaviour is seen in the non-linear constraints which entangle the central extensions, both among themselves in physical representations and, most strikingly, with the deformation appearing in the coproduct \cite{Gomez:2006va,Plefka:2006ze}.\footnote{{The complete symmetry algebra lying at the heart of the model was suspected to involve non-trivial deformations of ordinary quantum groups, possibly bordering with $W$-algebras \cite{Torrielli:2011zz}.}}  

Another exotic feature involves higher non-abelian symmetries \cite{Beisert:2007ds}. The Yangian charges are in correspondence with those at level 0, except for {the {\it secret} or {\it bonus} symmetry $\gen{B}$} \cite{Matsumoto:2007rh,Beisert:2007ty} which happens to have no correspondent. This generator is associated with the  hypercharge, {and thanks to an additional  {\it non-local tail} in its coproduct\footnote{{Beisert's $S$-matrix\cite{Beisert:2005tm} describes processes of the type $|\mbox{boson} \rangle \otimes |\mbox{boson} \rangle \mapsto |\mbox{fermion} \rangle \otimes |\mbox{fermion} \rangle$ and vice-versa, hence it does not preserve the hypercharge $(-)^F \otimes \mathbbm{1} + \mathbbm{1} \otimes (-)^F$.}} the associated hypercharge symmetry is recovered only at the first Yangian level.}
{For this reason it is not possible} to straightforwardly enlarge the symmetry algebra to the Yangian of $\alg{gl}(2|2)$. 

Progress was made \cite{Beisert:2014hya} relying on the $R\mathcal T\mathcal T$ formulation. {The authors of \cite{Beisert:2014hya} have shown that} starting from the the $S$-matrix in the fundamental representation, one can generate, via the $R\mathcal T\mathcal T$ relations, an abstract algebra capable of reproducing the salient ingredients of the $AdS_5$ integrable system, and where the {\it braiding element} is understood as a particular Yangian generator of level $-1$. Such indentation of levels had already been anticipated in the classical $r$-matrix algebra by \cite{Beisert:2007ty}, where Drinfeld's second realisation was explored \cite{Drinfeld:1987sy,Spill:2008tp}. 

Despite the successful embedding of the secret symmetry at the classical level in a consistent Lie-bialgebra picture \cite{Beisert:2007ty}, and the extremely promising $R\mathcal T\mathcal T$ formulation \cite{Beisert:2014hya}, a satisfactory universal form of the quantum-group symmetry, including its universal R-matrix, is still lacking. This is somehow reflected in the associated representation theory, for instance when {\it long} multiplets are studied \cite{Beisert:2006qh,Arutyunov:2009pw}. It has become progressively clearer \cite{Beisert:2017xqx} that extra generators, such as {\it e.g.} the full set of outer automorphisms, need to be added if one hopes to achieve a complete description. 

What we will demonstrate in this paper is that there is actually a missing part of the story, as we shall find {that an additional \emph{boost} generator is a symmetry of the $AdS_5/CFT_4$ string $S$-matrix, and it will eventually need to be consistently inserted into the final form of the algebra.}  
{Given that} the presence of the secret symmetry was remarkably shown not to be limited to the spectral problem---see the Conclusions---it would likewise be very exciting if this new boost symmetry should be investigated as well in other areas of ${\cal{N}}=4$ super Yang-Mills and of string theory on $AdS_5 \times S^5$.

\subsection{Quantum-deformed Poincar\'e supersymmetry}

{The boost generator is part} of a particular $q$-deformation of $1+1$-dimensional Poincar\'e superalgebra. We shall describe its commutation relations and coproduct action, discuss its classical limit and speculate on its possible universal interpretation. Before doing that, we need to set the stage by describing the relevant literature which brought us to consider this problem.
Let us {also point out that in the present context the super $q$-Poincar\'e deformation is an alternative interpretation of the algebraic structure of {\it ordinary} $AdS_5/CFT_4$, and it should be distinguished from other lines of investigation where the $q$-deformation is {\it superimposed}.\footnote{{These include the quantum-group deformations started in \cite{Beisert:2008tw,Klimcik:2008eq,Delduc:2014kha,Sfetsos:2013wia,Hollowood:2014qma}, see \cite{Borsato:2016hud} for a review. In \cite{Pachol:2015mfa} some of those results were related via a contraction to a deformed flat-space superstring with $q$-deformed Poincar\'e symmetry.
$q$-deformations have appeared also in \cite{Kawaguchi:2011wt,Kawaguchi:2012ug,Kawaguchi:2013lba} and in the Pohlmeyer-reduction of superstrings \cite{Hoare:2011fj}. } }
}
{In fact the idea---pursued in the papers that we are about to review and in the current one---is that fixing light-cone gauge on the $AdS_5 \times S^5$ string sigma-model does not really break the  Lorentz invariance on the worldsheet, but rather promotes it to a hidden symmetry, which is deformed as in the spirit of quantum groups.}

{In \cite{Gomez:2007zr}  the (massive) dispersion-relation of $AdS_5$ magnons was interpreted as coming from the Casimir }
\begin{equation}
C=\gen{H}^2+g^2 (\gen{K}^{\frac{1}{2}}-\gen{K}^{-\frac{1}{2}})^2, 
\end{equation}
of a $q$-deformation of the bosonic Poincar\'e algebra---see the following section---{containing also the boost generator  $\gen{J}$}  which shifts the torus-variable $z$
\begin{equation}
\gen{J}: z \to z + c.
\end{equation}
Immediately afterwards, the paper \cite{Young:2007wd} generalised this picture to encompass the entire centrally-extended $\alg{psu}(2|2)$ symmetry, by re-interpreting it as a $q$-deformed super-Poincar\'e algebra.\footnote{{We use the term ``super-Poincar\'e'' although this superalgebra possesses a non-standard feature, since the anticommutators of supercharges produce also the R-symmetry generators.}} {The coalgebra structure was also discussed, although only for a carefully chosen subalgebra of generators. The coproducts were fixed with the intention of reproducing the standard trigonometric quantum-group framework, faithful to the idea of \cite{Gomez:2007zr}, but this ultimately resulted in an incompatibility with the $S$-matrix. In fact, in that construction the energy generator ceases to be co-commutative, hence it fails to be a symmetry of the $S$-matrix.}

We shall demonstrate that one can overcome the difficulties with the coproduct by allowing a non-standard form, insisting on it being an $S$-matrix symmetry. 
We shall also argue that it is thanks to the shift in perspective of \cite{Gomez:2007zr,Young:2007wd}, namely putting emphasis on {\it quantum-group deformation of relativistic-type} symmetries, that one can extract a novel array of algebraic conditions constraining the scattering theory. It is also essential that the coupling constant $g$ is an {\it algebra} deformation parameter, rather than a {\it representation} one, an aspect which more closely fits the traditional approach to quantum integrable systems (cf. the Sine-Gordon theory).  
{We also find interesting connections with} the very recent \cite{Beisert:2017xqx}, on which we shall comment more diffusely in the Conclusions. 

In the spirit of \cite{Young:2007wd}, the paper \cite{Stromwall:2016dyw} demonstrated that an analogous $q$-Poincar\'e superalgebra could in fact be written down for {the sector of massless worldsheet excitations of the} $AdS_3$ superstrings. The $q$-deformed coproducts were fixed for all generators, and in this case the energy is an exact symmetry of the $S$-matrix {thanks to the massless dispersion relation}. The $q$-Casimir of this algebra again naturally reproduces the massless dispersion relation.\footnote{This new way of looking at the massless magnon supersymmetry enjoys a host of very natural traits, such as a compact reformulation of the comultiplication rule. Interesting connections with phonons and spinons, inspired by \cite{Ballesteros:1999ew}, became manifest in this setup. The boost was employed to derive a natural massless rapidity, reproducing Zamolodchikov's variable when taking the relativistic limit. This could be of help in providing insight from relativistic field theory into the proposals of \cite{Sax:2014mea,Tong:2014yna}.}
Following on, \cite{Fontanella:2016opq} {showed that in this construction the } boost-coproduct  is not a usual symmetry of the $AdS_3$ $S$-matrix, but rather it \emph{annihilates} it. The massless $S$-matrix in fact satisfies a system of differential equations whose closure singles out a flat connection, {which was then} used to re-express the $S$-matrix itself as a path-ordered exponential. {This lead to the proposal for a framework } to describe the $AdS_3$ massless sector in differential-geometric terms. 
We shall instead adopt here a different viewpoint from~\cite{Fontanella:2016opq}, as we are about to explain.

\subsection{This paper}

In this paper, we find an exact boost symmetry of the $AdS_5 \times S^5$ string $S$-matrix in the fundamental representation. This boost acts on single-particle {representations} as a derivative with respect to the momentum of the particle, dressed by a factor of the energy. We find the coproduct which makes this action a symmetry of the scattering matrix. The coproduct is both non-local and displays a non-trivial {\it tail}. We also analyse the antipode, related to crossing symmetry in physical representations, and discuss the constraint that this imposes on {the tail.}

{We study the semiclassical (strong-coupling) limit of the new symmetry we have found, and obtain its classical cobracket using the universal expression for the classical $r$-matrix given in \cite{Beisert:2007ty}. We rewrite the cobracket in a form which is suggestive of its quantum origin, and display the worldsheet realisation of the boost symmetry in terms of string sigma-model local currents.}

{We then study the opposite regime, namely the weak-coupling limit $g \to 0$, commenting upon the relationship to the spin-chain picture and on former literature on boosts in long-range lattice settings. We also consider the connection with the exceptional Lie superalgebra, in an attempt to establish a language which may allow to connect to the work of \cite{Beisert:2016qei,Beisert:2017xqx}. We finally write down a purely bosonic quantum affine Poincar\'e symmetry by employing a contraction of quantum affine algebra $U_q(\widehat{\alg{sl}_2})$, which, to our knowledge, is not available in the literature. This is in the spirit of trying to extend the boost symmetry to higher levels, see the Conclusions for further comments on this.}

\section{The boost and $q$-Poincar\'e symmetry}\label{sec:boost}
The integrable model under consideration is invariant under two copies of $\psu(2|2)_{c.e.}$~\cite{Beisert:2005tm}, where $c.e.$ means centrally extended. The three central elements---the hamiltonian $\gen{H}$ and two momentum-dependent charges $\gen{C},\overline{\gen{C}}$ conjugate to each other---are shared between the two copies. In the following we will only need to focus on one of such copies. 

The construction that follows the $\psu(2|2)_{c.e.}$ formulation is reviewed in Appendix~\ref{sec:rev-ads5}, where we also present our conventions. Here we follow Young~\cite{Young:2007wd} and we do not introduce the generators $\gen{C},\overline{\gen{C}}$, which are replaced already at the algebraic level by appropriate expressions of the momentum generator $\gen{P}$, or more conveniently its exponential $\gen{K}\equiv\exp(i\gen{P})$. 

 We will denote the 8 supercharges by $\gen{Q}_\alpha^{\ a},\, \overline{\gen{Q}}_a^{\ \alpha}$ where $a=1,2$ and $\alpha=3,4$ are indices corresponding to two copies\footnote{We need to identify $\gen{L}_1^{\ 1}=-\gen{L}_2^{\ 2}$ and $\gen{R}_3^{\ 3}=-\gen{R}_4^{\ 4}$.} of $\su(2)$ $\gen{L}_a^{\ b},\, \gen{R}_\alpha^{\ \beta}$, and bar denotes complex conjugation.
Together with $\gen{H},\gen{P}$ they span an ideal of the $q$-Poincar\'e superalgebra which we denote by $\alg{i}$.
The commutation relations involving the $\su(2)$ generators are\footnote{We use the convention $\epsilon_{12}=\epsilon^{12}=-\epsilon_{21}=-\epsilon^{21}=1$, $\epsilon_{34}=\epsilon^{34}=-\epsilon_{43}=-\epsilon^{43}=1$. When raising or lowering the indices we take $V^a=\epsilon^{ab}V_b,\ V_a=V^b\epsilon_{ba}$, and similarly for Greek indices.}
\be
\begin{aligned}
&[\gen{L}_a^{\ b},\gen{L}_c^{\ d}]=\delta^b_c\gen{L}_a^{\ d}-\delta^d_a\gen{L}_c^{\ b},\qquad
&&[\gen{R}_\alpha^{\ \beta},\gen{R}_\gamma^{\ \delta}]=\delta^\beta_\gamma\gen{R}_\alpha^{\ \delta}-\delta^\delta_\alpha\gen{R}_\gamma^{\ \beta},\\
&[\gen{L}_a^{\ b},\gen{q}_c]=\delta^b_c\gen{q}_a-\tfrac{1}{2}\delta_a^b\gen{q}_c,\qquad
&&[\gen{R}_\alpha^{\ \beta},\gen{q}_\gamma]=\delta^\beta_\gamma\gen{q}_\alpha-\tfrac{1}{2}\delta_\alpha^\beta\gen{q}_\gamma,\\
&[\gen{L}_a^{\ b},\gen{q}^c]=-\delta_a^c\gen{q}^b+\tfrac{1}{2}\delta_a^b\gen{q}^c,\qquad
&&[\gen{R}_\alpha^{\ \beta},\gen{q}^\gamma]=-\delta_\alpha^\gamma\gen{q}_\beta+\tfrac{1}{2}\delta_\alpha^\beta\gen{q}^\gamma,\\
\end{aligned}
\ee
where we used $\gen{q}$ to denote a generic element of $\alg{i}$. The anticommutators of supercharges give
\be\label{eq:comm-rel-noJ-AdS5}
\begin{aligned}
&\{\gen{Q}_\alpha^{\ a},\overline{\gen{Q}}_b^{\ \beta}\}= \delta^a_b \gen{R}_\alpha^{\ \beta}+\delta_\alpha^\beta\gen{L}_b^{\ a}+\tfrac{1}{2}\delta^a_b\delta_\alpha^\beta \gen{H},\\
&\{\gen{Q}_\alpha^{\ a},\gen{Q}_\beta^{\ b}\}=\tfrac{ig}{2}\, \epsilon_{\alpha\beta}\epsilon^{ab}\left(\gen{K}^{\frac{1}{2}}-\gen{K}^{-\frac{1}{2}}\right),\qquad
&&\{\overline{\gen{Q}}_a^{\ \alpha},\overline{\gen{Q}}_b^{\ \beta}\}=\tfrac{ig}{2}\, \epsilon^{\alpha\beta}\epsilon_{ab}\left(\gen{K}^{\frac{1}{2}}-\gen{K}^{-\frac{1}{2}}\right).
\end{aligned}
\ee
The above relations for the ideal $\alg{i}$ essentially reproduce those of $\psu(2|2)_{c.e.}$ in~\eqref{eq:comm-rel-psu22ce} if we formally identify\footnote{Comparing to the usual conventions in the literature, we preferred to redefine the supercharges in order to have $\gen{C}$ with a real eigenvalue.} $\gen{C}\sim\overline{\gen{C}}\sim \tfrac{ig}{2}\left(\gen{K}^{\frac{1}{2}}-\gen{K}^{-\frac{1}{2}}\right)$. In fact the two formulations are indistinguishable in the fundamental representation, see below.

The superalgebra considered here, however, contains an additional generator $\gen{J}$ that has no counterpart in the $\psu(2|2)_{c.e.}$ formulation. The generator $\gen{J}$ has the physical interpretation of a boost and it acts on $\alg{i}$ as\footnote{Our generators are related to those of~\cite{Celeghini:1990bf} as $\gen{P}=-iwP_x$, $\gen{H}=wgP_y$, $\gen{J}=igJ$.}~\cite{Young:2007wd}
\be\label{eq:comm-rel-J-AdS5}
\begin{aligned}
& [\gen{J},\gen{P}]=i\, \gen{H},\qquad\qquad
&&[\gen{J},\gen{Q}_\a^{\ a}]=-\tfrac{ig}{4}\left(\gen{K}^{\frac{1}{2}}+\gen{K}^{-\frac{1}{2}}\right) \epsilon_{\a\b}\epsilon^{ab}\overline{\gen{Q}}_b^{\ \b},\\
&[\gen{J},\gen{H}]=\tfrac{g^2}{2}\left(\gen{K}-\gen{K}^{-1}\right) ,\qquad
&&[\gen{J},\overline{\gen{Q}}_a^{\ \a}]=-\tfrac{ig}{4}\left(\gen{K}^{\frac{1}{2}}+\gen{K}^{-\frac{1}{2}}\right) \epsilon_{ab}\epsilon^{\a\b}\gen{Q}_\b^{\ b}.
\end{aligned}
\ee
It also follows $[\gen{J},\gen{K}]=-\gen{H}\gen{K}$. The two copies of $\su(2)$ commute with $\gen{J}$. Notice that when we include the boost, $\gen{H}$ and $\gen{P}$ cease to be central.

Let us emphasise that at the level of the  $q$-Poincar\'e subalgebra spanned by $\{\gen P,\gen H, \gen J\}$,  the parameter $g$---related as $g=\frac{\sqrt{\lambda}}{2\pi}$ to the 't Hooft coupling---is spurious since it can be removed by a rescaling of the generators. However, when we consider the full superalgebra this is no longer possible, and $g$ is a physical deformation parameter.\footnote{When obtaining the $q$-Poincar\'e bosonic algebra as a contraction of $U_q(\alg{sl}_2)$, one  takes a particular $q\to 1$ limit which leaves the algebra deformed~\cite{Celeghini:1990bf}. This also explains why we do not explicitly relate the deformation parameter $g$ to $q$.}

As already mentioned in the Introduction, the Casimir of the $q$-Poincar\'e subalgebra generated by $\gen{H},\gen{P},\gen{J}$ is $C=\gen{H}^2+g^2 (\gen{K}^{\frac{1}{2}}-\gen{K}^{-\frac{1}{2}})^2$ and setting the mass $C=1$ reproduces~\cite{Gomez:2007zr} the famous all-loop dispersion relation first conjectured in~\cite{Beisert:2004hm}.

\vspace{12pt}

In this paper we will often work with the fundamental representation of the above superalgebra~\cite{Beisert:2005tm}. We introduce states $\ket{\chi_a},\, a=1,2$ which are bosonic and $\ket{\chi_\alpha},\, \alpha=3,4$ which are fermionic. Then the $\su(2)$ generators act on them e.g. as 
$\gen{L}_a^{\ b}\ket{\chi_c}=\delta^b_c\ket{\chi_a}-\tfrac{1}{2}\delta_a^b\ket{\chi_c}$ and the supercharges as
\be
\begin{aligned}
&\gen{Q}_\a^{\ a}\ket{\chi_b}=a_p\ \delta^a_b\ket{\chi_\a},
\qquad
&&\overline{\gen{Q}}_{a}^{\ \a}\ket{\chi_b}=\bar b_p\ \epsilon_{ab}\epsilon^{\a\b}\ket{\chi_\b},\\
&\gen{Q}_\a^{\ a}\ket{\chi_\b}=b_p\ \epsilon_{\a\b}\epsilon^{ab}\ket{\chi_b},
\qquad
&&\overline{\gen{Q}}_{a}^{\ \a}\ket{\chi_\b}=\bar a_p\ \delta^{\a}_{\b}\ket{\chi_a}.
\end{aligned}
\ee
Here we have introduced parameters which depend on the coupling constant $g$ and the momentum $p$ labelling each excitation. We parameterise them as\footnote{This choice corresponds to $\xi=-p/4$ in (3.61) of~\cite{Arutyunov:2009ga}.}
\be\label{eq:par-repr}
a_p=\bar a_p=\sqrt{\frac{g}{2}}\gamma_p,\qquad
b_p=\bar b_p=-\sqrt{\frac{g}{2}}\frac{\gamma_p}{\sqrt{x^-_px^+_p}},
\qquad
\gamma_p=\sqrt{i(x^-_p-x^+_p)},
\ee
where the Zhukovski variables $x^\pm_p$ satisfy the constraints
\be
x^+_p+\frac{1}{x^+_p}-x^-_p-\frac{1}{x^-_p}=\frac{2i}{g},\qquad\quad
\frac{x^+_p}{x^-_p}=e^{ip}.
\ee
The states $\ket{\chi_n},\, n=1,\ldots,4$ may be represented in terms of the $n$-th unit vectors, so that the generators will be explicit $4\times 4$ matrices.
In the fundamental representation the central elements are clearly proportional to the identity matrix $\gen{H}=h_p\, \mathbf{1}, \ \gen{P}=p\, \mathbf{1}$ and one finds
\be
h_p=\frac{ig}{2}\left(x^--x^++\frac{1}{x^+}-\frac{1}{x^-}\right)=\sqrt{1+4g^2\sin^2\frac{p}{2}}.
\ee
In the fundamental representation the boost may be explicitly realised as $\gen{J}=ih_p\gen{1}\partial_p$,  it is easy to check that  this  reproduces all commutation relations in~\eqref{eq:comm-rel-J-AdS5}.
Let us mention that when taking the derivative with respect to $p$ we find convenient to implement it as
\be
\frac{\partial}{\partial p} = \frac{\partial x^-_p}{\partial p} \frac{\partial}{\partial x^-_p} +\frac{\partial x^+_p}{\partial p} \frac{\partial}{\partial x^+_p},
\ee
where
\be
\frac{\partial x^-_p}{\partial p}=\frac{i (x^-_p)^2 \left((x^+_p)^2-1\right)}{(x^-_p-x^+_p)
   (x^-_p x^+_p+1)},
\qquad
\frac{\partial x^+_p}{\partial p}= \frac{i \left((x^-_p)^2-1\right) (x^+_p)^2}{(x^-_p-x^+_p)
   (x^-_p x^+_p+1)}.
\ee
In terms of Janik's parameterisation of the rapidity torus the momentum and the energy are related to the rapidity $z$ as
\be
\sin\frac{p}{2}=\text{sn}(z,k),\qquad h_p=\text{dn}(z,k),\qquad k=-4g^2.
\ee
In this variable the boost is simply $\gen{J}=\tfrac{i}{2}\gen{1}\partial_z$.

\subsection{Coproduct}
In this section we equip the superalgebra with a coproduct to obtain the structure of a bialgebra.
To overcome the difficulties of~\cite{Young:2007wd}, i.e. the non-cocommutativity of $\gen H$ and the issues in  extending the coproduct to the full superalgebra as soon as we start including also the $\su(2)$ generators, we decide to follow a different route. For all generators that span the ideal $\alg{i}$ we keep the same coproduct that is used also in the $\psu(2|2)_{c.e.}$ formulation, see also Appendix~\ref{sec:rev-ads5}. In particular, we prefer to work in the {}{most symmetric frame}
\footnote{Although not explicitly written, the tensor product is graded. In the fundamental representation, when a supercharge appears in the second space of the tensor product, it should be accompanied by the action of the graded identity acting on the first space.}~\cite{Arutyunov:2009ga}
\be\label{eq:Delta-noJ}
\begin{aligned}
&\Delta({\gen{H}})=\gen{H}\otimes 1+1\otimes\gen{H}, \qquad
&&\Delta({\gen{P}})=\gen{P}\otimes 1+1\otimes\gen{P},\\
&\Delta({\gen{L}_a^{\ b}})=\gen{L}_a^{\ b}\otimes 1+1\otimes\gen{L}_a^{\ b},\qquad
&& \Delta({\gen{R}_\a^{\ \b}})=\gen{R}_\a^{\ \b}\otimes 1+1\otimes\gen{R}_\a^{\ \b},\\
&\Delta({\gen{Q}_\a^{\ a}})=\gen{Q}_\a^{\ a}\otimes \gen{K}^{-\frac{1}{4}}+\gen{K}^{\frac{1}{4}}\otimes\gen{Q}_\a^{\ a},\qquad
&&\Delta({\overline{\gen{Q}}_a^{\ \a}})=\overline{\gen{Q}}_a^{\ \a}\otimes \gen{K}^{\frac{1}{4}}+\gen{K}^{-\frac{1}{4}}\otimes\overline{\gen{Q}}_a^{\ \a}.
\end{aligned}
\ee
The coproduct for the momentum $\gen{P}$ is equivalent to writing $\Delta(\gen{K})=\gen{K}\otimes\gen{K}$.
We insist on having the above coproduct because we know that at least these generators will be symmetries for the $R$-matrix~\cite{Beisert:2005tm} that is obtained in the fundamental representation in the $\psu(2|2)_{c.e.}$ formulation
\be
\Delta^{op}(\gen{q})R=R\Delta(\gen{q}).
\ee
Here $^{op}$ is used to denote the opposite coproduct and we take $\Delta^{op}(\gen{q})=\Pi_g\cdot\Delta(\gen{q})\cdot\Pi_g$, where $\Pi_g$ is the graded permutation.\footnote{We assume that the first and second space in the tensor product always carry momenta ordered as $p_1,p_2$, even after applying $^{op}$. This means that in an explicit matrix realisation one should take care of explicitly swapping $p_1\leftrightarrow p_2$ after applying the graded permutation.} We refer to~\eqref{eq:S-mat} for the result of the $R$-matrix written in our conventions. 

The $R$-matrix is  compatible also with the generators of the Yangian constructed in~\cite{Beisert:2007ds} and the secret symmetry found in~\cite{Matsumoto:2007rh}, see Appendix~\ref{sec:rev-ads5}.
Compatibility with the above generators fixes the $R$-matrix only up to an overall scalar factor, and the normalisation in~\eqref{eq:S-mat} is arbitrary. The proper scalar factor is found upon solving the crossing equation~\cite{Janik:2006dc,Beisert:2006ez,Arutyunov:2009kf}, which is related to the possibility of introducing an antipode to promote the bialgebra to a Hopf algebra, {see also~\cite{Bernard:1992mu,Delius:1995he}. We refer to section~\ref{sec:antipode} for a discussion on the antipode of the boost.}

The $R$-matrix satisfies the Yang-Baxter equation (YBE), which is more conveniently written in term of the $S$-matrix $S=\Pi_g R$ as
\be
S_{12}(p_2,p_3)S_{23}(p_1,p_3)S_{12}(p_1,p_2)=S_{23}(p_1,p_2)S_{12}(p_1,p_3)S_{23}(p_2,p_3),
\ee
where subscripts denote the subspaces on which the $S$-matrix is acting, e.g. $S_{12}=S\otimes \gen 1$.

\vspace{12pt}

Let us finally discuss the possibility of determining a coproduct for $\gen{J}$ which makes it a symmetry of the $R$-matrix. We will do this just {}{in the fundamental representation}.

First, let us write $\Delta(\gen{J})=\Delta'(\gen{J})+\mathcal{T}$, where $\mathcal{T}$ is a tail that is assumed to commute with $\gen{H},\gen{P}$---all generators except $\gen{J}$ do so. Requiring that the coproduct is a homomorphism for the commutation relations of the $q$-Poincar\'e subalgebra span$\{\gen{H},\gen{P},\gen{J}\}$ we find
\be
\begin{aligned}
&\Delta'(\gen{J})=\left(1-\frac{ {s}_{12}}{h_1}\right)\gen{J}\otimes 1+\left(1+\frac{{s}_{12}}{h_2}\right)1\otimes \gen{J},\qquad
{s}_{12}=\frac{g}{2}\frac{\sin p_1+\sin p_2-\sin (p_1+p_2)}{w_1^{-1}-w_2^{-1}},
\end{aligned}
\ee
where we introduced
\be
\begin{aligned}
w_p&=\frac{2\, h_p}{g\ \sin p}=2\frac{1+x^-_px^+_p}{x^-_p+x^+_p}.
\end{aligned}
\ee
The tail $\mathcal T$ of the coproduct for the boost is obtained by demanding that the remaining commutation relations in~\eqref{eq:comm-rel-J-AdS5} are satisfied. It is not difficult to come up with an appropriate ansatz for $\mathcal{T}$ as a bilinear of (super)charges; commutation relations will then determine the unknown coefficients. We find that the tail is fixed up to a term $\mathcal{T}_1$ which in the fundamental representation is proportional to the identity matrix. The coproduct for the boost is\footnote{Notice that $\frac{1}{2}\left(1-\tan\frac{p_1}{2}\tan\frac{p_2}{2}\right)=\frac{x^-_1x^-_2+x^+_1 x^+_2}{(x^+_1+x^-_1)(x^+_2+x^-_2)} $.}
\begin{eqnarray}
&&\Delta(\gen{J})=\Delta'(\gen{J})+\mathcal{T}_{\gen{H}\hat{\gen{B}}}+\mathcal{T}_{\alg{psu}(2|2)}+\mathcal{T}_1,\label{eq:DeltaJ}\\ 
&& \nonumber \\
&&\mathcal{T}_{\gen{H}\hat{\gen{B}}}=\frac{1}{2} \frac{1}{w_1-w_2}\left(1-\tan\frac{p}{2}\otimes \tan\frac{p}{2}\right) \left(\gen{H}\otimes \hat{\gen{B}}+ \hat{\gen{B}}\otimes \gen{H}\right),\nonumber \\ 
&&\mathcal{T}_{\alg{psu}(2|2)}=\frac{1}{2}\frac{w_1+w_2}{w_1-w_2}\left(\gen{K}^{-\frac{1}{4}} \gen{Q}_\a^{\ a}\otimes \gen{K}^{-\frac{1}{4}} \overline{\gen{Q}}_a^{\ \a}
-\gen{K}^{\frac{1}{4}} \overline{\gen{Q}}_a^{\ \a}\otimes \gen{K}^{\frac{1}{4}} \gen{Q}_\a^{\ a}
+\gen{L}_a^{\ b}\otimes\gen{L}_{ b}^{\ a}
-\gen{R}_\a^{\ \b}\otimes\gen{R}_{ \b}^{\ \a}\right).\nonumber
\end{eqnarray}
Interestingly, to write the result we need  the secret symmetry $\hat{\gen{B}}$, see~\eqref{eq:secret}. We refer to Section~\ref{sec:cobracket} for an argument that  motivates why this should be the case.

As mentioned, the element $\mathcal{T}_1$ is not fixed by this computation. Moreover, at this point we do not know yet whether $\Delta(\gen{J})$ is a symmetry of the $R$-matrix. It turns out that {}{(the antisymmetric part of)} $\mathcal{T}_1$ is fixed precisely by demanding that the boost is a symmetry of $R$.  In fact, we can compute\footnote{In principle one  acts with the derivative also on ``test states'' on which these operators are acting. However, after using Leibniz's rule the terms with derivatives on the states cancel. Notice that by the same argument the usual boost-covariance of the relativistic S-matrix $(\partial_{\theta_1}+\partial_{\theta_2})S(\theta_1,\theta_2)=0$ may also be interpreted as the requirement of the vanishing of the commutator of the boost (here represented by $\partial_\theta$) with $S$. We thank Niklas Beisert for pointing this out.}
\be
\begin{aligned}
&\Delta^{op}(\gen{J})R-R\Delta(\gen{J})=
i\left[(h_1- {s}_{12})\partial_{p_1}+(h_2+{s}_{12})\partial_{p_2}\right]R
+\mathcal{T}^{op}R-R\mathcal{T},
\end{aligned}
\ee
and we can check explicitly that the above gives
\be\label{eq:centr-rem}
\begin{aligned}
&\Delta^{op}(\gen{J})R-R\Delta(\gen{J})=(f_{12}+\mathcal{T}_1^{op}-\mathcal{T}_1)\, R,\\
&f_{12}=-\frac{1}{2}\frac{w_1+w_2}{w_1-w_2}-
\frac{ig}{2} \frac{x^-_1x^-_2+x^+_1 x^+_2}{(x^+_1+x^-_1)(x^+_2+x^-_2)}
\frac{(x^-_1-1/x^-_1)(x^-_2-1/x^-_2)-(x^+_1-1/x^+_1)(x^+_2-1/x^+_2)}
{w_1-w_2}.
\end{aligned}
\ee
The  non-trivial point of this result is that the remainder of this equation is proportional to the $R$-matrix itself. We therefore conclude that it is indeed possible to cancel the unwanted terms by 
 fixing $\mathcal{T}_1=f_{12}/2+\mathcal T^s_1$. Here $\mathcal T^s_1$ is the part symmetric under $^{op}$. Boost invariance does not constrain it since it drops out of the equation.

\vspace{12pt}

Obviously, because of the presence of the derivative, the above computation is sensitive to the normalisation used for the $R$-matrix. It is also clear that a different normalisation for the $R$-matrix will just produce new terms in the equation that are still proportional to $R$ itself, since they will just come from acting with the derivatives on the new scalar factor. It will be therefore enough to further shift $\mathcal{T}_1$ by {}{an appropriate counterterm} in order to make sure that the boost is still a symmetry. In other words, if we define $R'=e^{\Phi_{12}}R$ we then get
\be\label{eq:compat-new-norm}
\Delta^{op}(\gen{J})R'-R'\Delta(\gen{J})=[f_{12}{+\mathcal{T}_1^{op}-\mathcal{T}_1}+i(h_1- {s}_{12})\partial_{p_1}\Phi_{12}+i(h_2+{s}_{12})\partial_{p_2}\Phi_{12}]R'.
\ee
{Taking $\mathcal{T}_1=\frac{1}{2} [f_{12}+i(h_1- {s}_{12})\partial_{p_1}\Phi_{12}+i(h_2+{s}_{12})\partial_{p_2}\Phi_{12}]\mathbf{1}{+\mathcal T_1^s}$ would ensure} that  $\gen{J}$ is  a symmetry of  $R'$. 
These  considerations put the boost on a special footing when compared to the other charges. In fact, if the coproduct $\Delta(\gen{J})$ were {\emph{a priori}} known{---including the tail $\mathcal T_1$---}demanding that it is a symmetry of $R$ would produce a constraint on its scalar factor in the form of a partial differential equation.
{On the other hand, here we are going in the opposite direction---we first fix the scalar factor and from that we determine the tail $\mathcal T_1$---so that any choice for $\Phi_{12}$ (or $\mathcal T_1$) seems legitimate. We will see in the next subsection that  
the symmetric part of the tail $\mathcal T_1^s$ will have to satisfy certain constraints in order to be consistent with the introduction of an antipode for $\gen J$, which is necessary to place the boost into the framework of Hopf algebras. This is indeed not surprising, since we know that crossing symmetry constrains the allowed scalar factor~\cite{Bernard:1992mu,Delius:1995he,Janik:2006dc}, and must therefore restrict also the allowed $\mathcal T_1$.}

It is natural at this point to consider the contribution of the BES phase~\cite{Beisert:2006ez}. This is given by $\theta(p_1,p_2)=\chi(x^+_1,x^+_2)+\chi(x^-_1,x^-_2)-\chi(x^+_1,x^-_2)-\chi(x^-_1,x^+_2)$ where~\cite{Dorey:2007xn}
\be
\begin{aligned}
\chi(x_1,x_2)&=i\ \oint\frac{dv_1}{2\pi i} \oint\frac{dv_2}{2\pi i}\ \frac{1}{x_1-v_1} \frac{1}{x_2-v_2}\ G(v_1,v_2),
\\
G(v_1,v_2)&=\log\frac{\Gamma[1+i\tfrac{h}{2}(v_1+1/v_1-v_2-1/v_2)]}{\Gamma[1-i\tfrac{h}{2}(v_1+1/v_1-v_2-1/v_2)]}.
\end{aligned}
\ee
If we define for simplicity the operator $\mathbb{D}\equiv i(h_1- {s}_{12})\partial_{p_1}+i(h_2+{s}_{12})\partial_{p_2}$, we have\footnote{Alternatively, one may use the representation of~\cite{Volin:2009uv} for the BES phase, which is equivalent to the DHM representation~\cite{Dorey:2007xn} used here. We prefer the DHM representation since the derivative will act only on rational expressions of the variables $x^\pm_i$. In the representation of~\cite{Volin:2009uv} the momentum-dependence is also through the Gamma functions.}
\be\label{eq:contr-BES-tail}
\begin{aligned}
&\mathbb{D}[\theta(p_1,p_2)]=i\ \oint\frac{dv_1}{2\pi i} \oint\frac{dv_2}{2\pi i}\ K(x^\pm_1,v_1,x^\pm_2,v_2)\ G(v_1,v_2),\\
&K(x^\pm_1,v_1,x^\pm_2,v_2)\equiv\mathbb{D}\left[ \left(\frac{1}{x^+_1-v_1}-\frac{1}{x^-_1-v_1}\right)\left(\frac{1}{x^+_2-v_2}-\frac{1}{x^-_2-v_2}\right)\right].
\end{aligned}
\ee
The expression $K(x^\pm_1,v_1,x^\pm_2,v_2)$ is straightforwardly computed, and we omit it. It would be interesting to see how far it is possible to go when  performing the explicit integral.

\subsection{Antipode}\label{sec:antipode}
{A} bialgebra\footnote{The check of coassociativity of $\Delta$ when including $\gen J$ is challenging; the difficulty is related to the presence of higher-level generators in the tail of $\Delta(\gen J)$ (see also section~\ref{sec:cobracket}) and to the lack of a universal formulation. We thank Marius de Leeuw for comments on this.} may be promoted to a Hopf algebra if it is possible to introduce  an antipode $S$ that satisfies
\be\label{eq:axiom-antipode}
\mu \circ(S\otimes \text{id})\circ\Delta=\mathbf{1}\circ \epsilon=\mu \circ( \text{id}\otimes S)\circ\Delta
\ee
where $\epsilon$ is the counit, id the unit, $\Delta$ the coproduct and $\mu$ the multiplication.\footnote{{Here} we can take $\epsilon(\gen{1})=1$, and $\epsilon(\gen{q})=0$ for all other generators $\gen{q}$.}
In the {}{most symmetric frame}, on all generators in the ideal $\alg{i}$ the antipode acts as $S(\gen{q})=-\gen{q}$. We will assume that the antipode acts also on the secret symmetry in the same fashion.\footnote{In~\cite{Matsumoto:2007rh} the antipode of the secret symmetry was found to receive an additional shift $S(\hat{\gen{B}})=-\hat{\gen{B}}+c$ which is proportional to the identity matrix in the fundamental representation. However, it is enough to add a central piece to the tail of its coproduct in~\eqref{eq:DeltaB} in order to cancel the shift in the antipode. We will assume this here, so that the antipode can be implemented by~\eqref{eq:antipode-repr} also on $\hat{\gen{B}}$.}
In the fundamental representation the antipode of a generic charge is obtained as
\be\label{eq:antipode-repr}
S(\gen{q}(p))=\mathscr C\, \gen{q}^{st}(\bar p)\, \mathscr C^{-1},
\qquad\qquad
\mathscr C=\left(
\begin{array}{cccc}
 0 & -i & 0 & 0 \\
 i & 0 & 0 & 0 \\
 0 & 0 & 0 & 1 \\
 0 & 0 & -1 & 0 \\
\end{array}
\right),
\ee
where $st$ denotes supertransposition and $\mathscr C$ is the charge conjugation matrix. The crossed momentum $\bar p$ is obtained by sending $x^\pm_p\to 1/x^\pm_p$ everywhere, except for $\gamma_p$ in~\eqref{eq:par-repr} which should be analytically continued with more care~\cite{Janik:2006dc,Arutyunov:2009ga}
\be
\gamma_p\to
-\frac{i}{x^+_p}\left(\frac{x^+_p}{x^-_p}\right)^{1/2}\gamma_p.
\ee

\vspace{12pt}

In order to find how the antipode acts on the boost, we consider equation~\eqref{eq:axiom-antipode} assuming that the counit acts trivially on $\gen{J}$ as well. We therefore fix $S(\gen{J})$ by solving the equation
\be\label{eq:mu-s-delta}
\mu \circ(S\otimes \text{id})\circ\Delta(\gen{J})=0.
\ee
Since we do not have a universal expression for $\Delta(\gen{J})$ we will do this computation in the fundamental representation only.
When applying the multiplication $\mu$ we will first identify the two vector spaces in the tensor product, and susequently take the limit $p_2\to p_1$. The limiting procedure is crucial as we will show in a moment. Let us study separately the various contributions coming from $\Delta(\gen{J})$ in~\eqref{eq:DeltaJ}. The first contribution is 
\be\label{eq:muSDelta-prime}
\mu \circ(S\otimes \text{id})\circ\Delta'(\gen{J})=\left(1+\frac{\ell_p}{h_p}\right)(S(\gen{J})+\gen{J}),
\qquad
\ell(p_1)\equiv \lim_{p_2\to p_1} \, {s}(\bar p_1,p_2),
\ee
where a direct computation gives
\be
\ell_p=\frac{g}{2}\, w^2_p\left(\frac{dw_p}{dp}\right)^{-1}(\cos p-1).
\ee
Turning now to the contributions coming from the tail of the coproduct, we immediately see that for each of them there is a potential pole in the $p_2\to p_1$ limit coming from $(w_1-w_2)^{-1}$. As a first step we will look at the residues of these poles and check that they cancel each other, to make sure that we get a consistent result. We find
\be
\begin{aligned}
\mu \circ(S\otimes \text{id})\circ \mathcal{T}_{\gen{H}\hat{\gen{B}}}&=
\lim_{p_2\to p_1} \,\frac{1}{w_1-w_2}\left(-h_1 \hat{b}_1 (1+\tan^2\frac{p_1}{2})\right)\mathbf{1}_g,\\
\mu \circ(S\otimes \text{id})\circ \mathcal{T}_{\psu(2|2)}&=
\lim_{p_2\to p_1} \,\frac{-w_1}{w_1-w_2}
\left(\gen{Q}_\a^{\ a} \overline{\gen{Q}}_a^{\ \a}-\overline{\gen{Q}}_a^{\ \a} \gen{Q}_\a^{\ a}
+\gen{L}_a^{\ b}\gen{L}_{ b}^{\ a}-\gen{R}_\a^{\ \b}\gen{R}_{ \b}^{\ \a}\right)+\text{finite}\\
&=\lim_{p_2\to p_1} \,\frac{1}{w_1-w_2}\, \frac{w_1}{2}\mathbf{1}_g+\text{finite},
\end{aligned}
\ee
where we have denoted by $\hat{b}_p$ the eigenvalue of the secret symmetry $\hat{\gen{B}}=\hat{b}_p\mathbf{1}_g$. It is straightforward to check that
$w_p=2h_p\hat{b}_p(1+\tan^2\frac{p}{2})$, so that the divergences cancel.

The piece with the supercharges actually gives also a finite contribution, since the braiding factors $\gen{K}^{\pm \frac{1}{4}}$ appearing there are responsible for a term of order $(p_1-p_2)$ which multiplies $(w_1-w_2)^{-1}$. If we write $p_2=p_1+\epsilon$
\be
\begin{aligned}
\mu \circ(S\otimes \text{id})\circ \mathcal{T}_{\psu(2|2)}&=
\lim_{\epsilon\to 0} \,\frac{1}{w(p_1+\epsilon)/w(p_1)-1}
\left(e^{-i\epsilon/4}\gen{Q}_\a^{\ a} \overline{\gen{Q}}_a^{\ \a}-e^{i\epsilon/4}\overline{\gen{Q}}_a^{\ \a} \gen{Q}_\a^{\ a}
\right)+\text{pole}\\
&=d_{p_1}\gen 1+\text{pole},
\qquad\qquad
d_p\equiv-\frac{iw_p}{2}\left(\frac{dw_p}{dp}\right)^{-1}h_p.
\end{aligned}
\ee
We should not forget the contribution coming from the  tail $\mathcal{T}_1$, and if we define $c_p^{[1]}\gen{1}\equiv \mu \circ(S\otimes \text{id})\circ\mathcal{T}_1$ and solve~\eqref{eq:mu-s-delta} we finally get
\be
S(\gen{J})=-\gen{J}-\left(1+\frac{\ell_p}{h_p}\right)^{-1}\left(d_p+c^{[1]}_p\right)\gen{1}.
\ee
Therefore, in general the antipode on $\gen J$ will not just reverse its sign, but it could also introduce a shift proportional to $\gen 1$. We will soon be more explicit and work out the contribution $c^{[1]}_p$.

An important remark is that so far we have determined $S(\gen J)$ by solving~\eqref{eq:mu-s-delta}, but we could have instead looked at $\mu \circ( \text{id}\otimes S)\circ\Delta(\gen{J})=0$, which should hold as well. The calculations procede as above, in particular the contribution of $\Delta'(\gen J)$ is still given by~\eqref{eq:muSDelta-prime} and all apparent poles still cancel. However, when we apply the antipode on the second space  the  finite contribution $d_p$ appears with the \emph{opposite} sign. 

In other words we  obtain 
\be
S(\gen{J})=-\gen{J}-\left(1+\frac{\ell_p}{h_p}\right)^{-1}\left(-d_p+c^{[2]}_p\right)\gen{1},
\ee
where we defined $c_p^{[2]}\gen{1}\equiv \mu \circ( \text{id}\otimes S)\circ\mathcal{T}_1$. The two results for the antipode $S(\gen{J})$ are compatible only if 
\be\label{eq:compat-antipode}
d_p+c^{[1]}_p=-d_p+c^{[2]}_p.
\ee
As we will see in a moment this puts constraints on the symmetric part of the tail $\mathcal T_1^s$.
Let us define 
\be
\begin{aligned}
c^{[1,f]}_p &\equiv \mu \circ(S\otimes \text{id})\circ (\tfrac{1}{2} f_{12}\gen 1),\\
c^{[1,\Phi]}_p &\equiv \mu \circ(S\otimes \text{id})\circ [ \tfrac{i}{2}(h_1- {s}_{12})\partial_{p_1}\Phi_{12}+\tfrac{i}{2}(h_2+{s}_{12})\partial_{p_2}\Phi_{12}]\mathbf{1},\\
c^{[1,s]}_p &\equiv \mu \circ(S\otimes \text{id})\circ \mathcal T_1^s,
\end{aligned}
\ee
and similarly for the quantities where we act with $(\text{id}\otimes S)$ instead, so that $c^{[i]}_p=c^{[i,f]}_p+c^{[i,\Phi]}_p+c^{[i,s]}_p, i=1,2$.
Here we include the contributions $c^{[i,\Phi]}_p$ since we are considering the $R$-matrix $R'=e^{\Phi_{12}}R$, and we require boost invariance of $R'$---see the discussion around~\eqref{eq:compat-new-norm}.
In fact, it is well known~\cite{Janik:2006dc} that it is necessary to introduce a scalar factor $e^{\Phi_{12}}$ in order to have an $R$-matrix compatible with the antipode as $(S\otimes 1)R' = (R')^{-1}$. This crossing symmetry of the $R$-matrix\footnote{More explicitly it reads as $(\mathscr C\otimes 1)(R'(\barup{p}_1,p_2))^{st_1}(\mathscr C^{-1}\otimes 1)=(R'(p_1,p_2))^{-1}$.} is equivalent to the following equation for the phase
\be
\Phi_{\barup{1} 2}+\Phi_{12} = g_{12},\qquad
g_{12} = \log\left(\frac{x^-_1-x^-_2}{x^+_1-x^-_2}\ \frac{1-\frac{1}{x^-_1x^+_2}}{1-\frac{1}{x^+_1x^+_2}}\right).
\ee
Here when  crossing  the momentum $p_1$ we use the notation $\barup 1$ to remind that crossing is implemented by shifting \emph{upwards} by half of the period of the rapidity torus, see e.g.~\cite{Arutyunov:2009ga}. Shifting \emph{downwards} is denoted by $\bardown p$ and it is necessary when writing the crossing equation in the second variable $\Phi_{1\bardown{2}}+\Phi_{12} =- g_{\bardown{2}1}$. In both cases $\barup p=-p=\bardown p$.
We will now show that the quantities $c^{[1,\Phi]}_p,c^{[2,\Phi]}_p$ can be calculated just from the crossing equation for $\Phi_{12}$. This way of proceeding is simpler than studying the analytic continuation of~\eqref{eq:contr-BES-tail} to the crossed region, as it was done in~\cite{Arutyunov:2009kf} for the phase itself. From their definitions we first obtain
\be
\begin{aligned}
c^{[1,\Phi]} = \tfrac{i}{2}(h_1+\ell_1)\lim_{p_2\to p_1}\left(\Phi^{(2)}_{\barup 1 2}-\Phi^{(1)}_{\barup 1 2}\right),\\
c^{[2,\Phi]} = \tfrac{i}{2}(h_1+\ell_1)\lim_{p_2\to p_1}\left(\Phi^{(1)}_{1 \bardown 2}-\Phi^{(2)}_{1 \bardown 2}\right),
\end{aligned}
\ee
where superscripts $(1)$ and $(2)$ are used to denote derivative with respect to the first or second argument.\footnote{Notice that for example $\Phi^{(2)}_{21} = \partial_{p_1}\Phi(p_2,p_1)$ and that $\Phi^{(1)}_{\barup 1 2} = \partial_{\barup p_1}\Phi_{\barup 1 2}=-\partial_{p_1}\Phi_{\barup 1 2}$.}
From the derivatives of the crossing equation $-\Phi^{(1)}_{\barup 1  2}+\Phi^{(1)}_{12} = g^{(1)}_{12}$ and $\Phi^{(2)}_{\barup 1 2}+\Phi^{(2)}_{12} = g^{(2)}_{12}$ we obtain
\be
\begin{aligned}
c^{[1,\Phi]} = \tfrac{i}{2}(h_1+\ell_1)\lim_{p_2\to p_1}\left(-\Phi^{(2)}_{1 2}+g^{(2)}_{1 2}-\Phi^{(1)}_{ 1 2}+g^{(1)}_{12}\right)= \tfrac{i}{2}(h_1+\ell_1)\lim_{p_2\to p_1}\left(g^{(2)}_{1 2}+g^{(1)}_{12}\right).
\end{aligned}
\ee
In the last step the contributions coming from the phase cancel each other, because inside the limit we can interchange the two momenta and we can use the unitarity relation $\Phi_{21}=-\Phi_{12}$. With similar considerations one finds
\be
c^{[2,\Phi]} =- \tfrac{i}{2}(h_1+\ell_1)\lim_{p_2\to p_1}\left(g^{(2)}_{ \bardown 1 2}-g^{(1)}_{ \bardown 1 2}\right).
\ee
It is easy to see that\footnote{Explicitly we have $c^{[1,\Phi]}-c^{[2,\Phi]}=\tfrac{i}{2}(h_1+\ell_1)\lim_{p_2\to p_1}\left(\partial_{p_2}g_{12}+\partial_{p_1}g_{12}+\partial_{p_2}g_{\bar 12}+\partial_{p_1}g_{\bar 12}\right)
=\tfrac{i}{2}(h_1+\ell_1)\lim_{p_2\to p_1}\partial_{p_2}\left(g_{12}+g_{21}+g_{\bar 12}+g_{\bar 21}\right)
=\tfrac{i}{2}(h_1+\ell_1)\lim_{p_2\to p_1}\partial_{p_2}\log 1=0$, where we first relabelled $p_1\leftrightarrow p_2$ and then used the shortening condition for $x^\pm_i$. Here it is not necessary to specify in which direction we shift the rapidity variable.} $c^{[1,\Phi]}=c^{[2,\Phi]}$.
To conclude, the compatibility with the antipode written in~\eqref{eq:compat-antipode} is equivalent to
\be
c^{[2,s]}-c^{[1,s]} = 2d_p +c^{[1,f]}-c^{[2,f]} +c^{[1,\Phi]}-c^{[2,\Phi]}=2d_p.
\ee
In the last step we have used also $c^{[1,f]}_p=c^{[2,f]}_p=0$ which is easily checked starting from their definitions. 
As anticipated, the boost is compatible with crossing symmetry only if the symmetric tail $\mathcal T_1^s$ (which cannot be fixed by boost invariance) is such that the above equation is satisfied. Since it implies $c^{[2,s]}\neq c^{[1,s]}$, we deduce that it matters if we are shifting upwards or downwards when implementing crossing, in other words $\mathcal T_1^s$ cannot be written as a meromorphic function of $x^\pm_i$.

{We conclude that in our setup boost invariance cannot be used to put additional constraints on the dressing phase, as the coproduct tail can always be reverse-engineered to accomodate any dressing factor satisying  crossing.}
Let us remark that these results are quite general and have been obtained just by using the fact that $\Phi_{12}$ solves the crossing equation. Therefore, the results include also the case in which the $R$-matrix is normalised with the physical BES phase. As expected, CDD factors would not modify the above discussion, since by definition they solve the homogeneous crossing equation.

\section{Semiclassical limit}\label{sec:semcl}
In this section we start by reviewing some facts regarding the semiclassical limit of the $q$-Poincar\'e superalgebra under study. 
The semiclassical limit is achieved by first rescaling 
\be\label{eq:rescale-JP}
\gen{J}\to g\ \gen{J},
\qquad\qquad
\gen{P}\to \gen{P}/g,
\ee
and then taking $g\to\infty$. This is essentially the BMN limit~\cite{Berenstein:2002jq} in the $\psu(2|2)_{c.e.}$ formulation. Notice that in that case the semiclassical limit is a contraction at the level of the {}{representation}, while here it is a contraction at the level of the {}{algebra}. In this limit the deformation is lost and one obtains a classical Lie superalgebra. For the ideal $\alg{i}$ one finds
\be\label{eq:comm-rel-AdS5-BMN}
\begin{aligned}
&\{\gen{Q}_\alpha^{\ a},\overline{\gen{Q}}_b^{\ \beta}\}= \delta^a_b \gen{R}_\alpha^{\ \beta}+\delta_\alpha^\beta\gen{L}_b^{\ a}+\tfrac{1}{2}\delta^a_b\delta_\alpha^\beta \gen{H},\\
&\{\gen{Q}_\alpha^{\ a},\gen{Q}_\beta^{\ b}\}=-\tfrac{1}{2}\epsilon_{\alpha\beta}\epsilon^{ab}\gen{P},\qquad
&&\{\overline{\gen{Q}}_a^{\ \alpha},\overline{\gen{Q}}_b^{\ \beta}\}=-\tfrac{1}{2}\epsilon^{\alpha\beta}\epsilon_{ab}\gen{P},\\
&[\gen{L}_a^{\ b},\gen{L}_c^{\ d}]=\delta^b_c\gen{L}_a^{\ d}-\delta^d_a\gen{L}_c^{\ b},\qquad
&&[\gen{R}_\alpha^{\ \beta},\gen{R}_\gamma^{\ \delta}]=\delta^\beta_\gamma\gen{R}_\alpha^{\ \delta}-\delta^\delta_\alpha\gen{R}_\gamma^{\ \beta},\\
&[\gen{L}_a^{\ b},\gen{q}_c]=\delta^b_c\gen{q}_a-\tfrac{1}{2}\delta_a^b\gen{q}_c,\qquad
&&[\gen{R}_\alpha^{\ \beta},\gen{q}_\gamma]=\delta^\beta_\gamma\gen{q}_\alpha-\tfrac{1}{2}\delta_\alpha^\beta\gen{q}_\gamma,\\
&[\gen{L}_a^{\ b},\gen{q}^c]=-\delta_a^c\gen{q}^b+\tfrac{1}{2}\delta_a^b\gen{q}^c,\qquad
&&[\gen{R}_\alpha^{\ \beta},\gen{q}^\gamma]=-\delta_\alpha^\gamma\gen{q}_\beta+\tfrac{1}{2}\delta_\alpha^\beta\gen{q}^\gamma.\\
\end{aligned}
\ee
The only relations that are  affected by the  limit are those that involve  the momentum $\gen{P}$. 
In the semiclassical limit the boost  acts on the generators of $\alg{i}$ as
\be\label{eq:comm-rel-J-AdS5-BMN}
\begin{aligned}
& [\gen{J},\gen{P}]=i\, \gen{H},\qquad\qquad
&&[\gen{J},\gen{Q}_\a^{\ a}]=-\tfrac{i}{2}\, \epsilon_{\a\b}\epsilon^{ab}\overline{\gen{Q}}_b^{\ \b},\\
&[\gen{J},\gen{H}]=i\, \gen{P} ,\qquad
&&[\gen{J},\overline{\gen{Q}}_a^{\ \a}]=-\tfrac{i}{2}\, \epsilon_{ab}\epsilon^{\a\b}\gen{Q}_\b^{\ b},
\end{aligned}
\ee
and we recognise a Poincar\'e subalgebra spanned by $\{\gen{H},\gen{P},\gen{J}\}$. The Casimir now reduces to $C=\gen{H}^2+ \gen{P}^2$ which gives the usual relativistic dispersion relation.
Interestingly, in the semiclassical limit the boost may be identified with the combination of $\alg{sl}_2$ generators $\gen{J}\sim -\frac{i}{2}(\gen{B}_++\gen{B}_-)$ in~\eqref{eq:outer-sl2}.

In the semiclassical limit all the coproducts, including the one for the boost, reduce to the trivial ones
\be\label{eq:Delta-triv}
\begin{aligned}
&\Delta({\gen{H}})=\gen{H}\otimes 1+1\otimes\gen{H}, \qquad
&&\Delta({\gen{P}})=\gen{P}\otimes 1+1\otimes\gen{P},\\
&\Delta({\gen{L}_a^{\ b}})=\gen{L}_a^{\ b}\otimes 1+1\otimes\gen{L}_a^{\ b},\qquad
&& \Delta({\gen{R}_\a^{\ \b}})=\gen{R}_\a^{\ \b}\otimes 1+1\otimes\gen{R}_\a^{\ \b},\\
&\Delta({\gen{Q}_\a^{\ a}})=\gen{Q}_\a^{\ a}\otimes 1+1\otimes\gen{Q}_\a^{\ a},\qquad
&&\Delta({\overline{\gen{Q}}_a^{\ \a}})=\overline{\gen{Q}}_a^{\ \a}\otimes 1+1\otimes\overline{\gen{Q}}_a^{\ \a},\\
&\Delta({\gen{J}})=\gen{J}\otimes 1+1\otimes\gen{J}.
\end{aligned}
\ee
The fact that also the boost has a trivial coproduct in the semiclassical limit suggests that it should be realised as a local charge on the {}{worldsheet}. We will confirm this in section~\ref{sec:ws}

\vspace{12pt}

As noticed in~\cite{Young:2007wd} the above classical superalgebra may be obtained as a contraction of $\alg{d}(2,1,\varepsilon)$.  
This superalgebra contains 8 supercharges that we denote by $\gen{S}^{a\alpha \hat{a}}$; they carry 3 indices since they transform in the fundamental representation of 3 copies of $\su(2)$---whose commutation relations we do not repeat. The anticommutator of two supercharges reads as
\be
\{\gen{S}^{a\alpha \hat{a}},\gen{S}^{b\beta\hat{b}}\}=(1-\varepsilon)\epsilon^{\a\b}\epsilon^{\hat{a}\hat b}\gen{L}^{ab}
-\epsilon^{ab}\epsilon^{\hat{a}\hat b}\gen{R}^{\a\b}
+\varepsilon\  \epsilon^{ab}\epsilon^{\a\b}\gen{J}^{\hat a\hat b}.
\ee
If we identify the generators as\footnote{The following identification differs from the one of~\cite{Young:2007wd} by an {}{appropriate} $\su(2)$ rotation, since we want to make sure that the momentum $\gen{P}$ is identified with the Cartan of $\su(2)_{\gen{J}}$.} (we take $\hat a=5,6$)
\be\label{eq:cl-d21a-sP}
\begin{aligned}
&\gen{Q}^{\a a}=\tfrac{1}{\sqrt{2}}\left(\gen{S}^{a\alpha {5}}-i\gen{S}^{a\alpha {6}}\right),\qquad
&&\gen{H}=i\varepsilon\left(\gen{J}_{ 6}^{\  5}-\gen{J}_{ 5}^{\  6}\right),\qquad
&&&\gen{J}=-\frac{i}{2}\left(\gen{J}_{ 5}^{\  6}+\gen{J}_{ 6}^{\  5}\right),\\
&\overline{\gen{Q}}^{a\a}=\tfrac{1}{i\sqrt{2}}\left(\gen{S}^{a\alpha {5}}+i\gen{S}^{a\alpha {6}}\right),\qquad
&&\gen{P}=2i\varepsilon\gen{J}_{ 6}^{\  6},
\end{aligned}
\ee
then we easily check that in the  $\varepsilon\to 0$ limit we reproduce the (anti)commutation relations~\eqref{eq:comm-rel-AdS5-BMN},\eqref{eq:comm-rel-J-AdS5-BMN}.

Let us remind that  $\psu(2|2)_{c.e.}$ may be recovered from a different contraction of $\alg{d}(2,1,\varepsilon)$~\cite{Beisert:2005tm}, where also the combination $\gen{J}_{ 5}^{\  6}+\gen{J}_{ 6}^{\  5}$  scales as $\varepsilon^{-1}$---all three generators of $\su(2)_{\gen{J}}$ become central, and can be identified with $\gen{H},\gen{C},\overline{\gen{C}}$.

\subsection{The boost on the worldsheet}\label{sec:ws}
The $q$-Poincar\'e superalgebra  should correspond to the symmetries of superstrings on $AdS_5\times S^5$ in light-cone gauge, when the on-shell condition---$p=0$ in the absence of winding \cite{Arutyunov:2009ga}---is relaxed. After fixing light-cone gauge relativistic invariance on the worldsheet is lost, and our findings on $\Delta(\gen{J})$ in~\eqref{eq:DeltaJ} suggest that the  boost  will be a non-local symmetry of the model. Relativistic invariance is recovered in the semiclassical limit, and in fact in this limit the boost can be written as a local charge on the worldsheet.

In the large-tension limit  the Hamiltonian on the worldsheet is that of 8 free bosons--- 4 from $AdS_5$ denoted by $Z^{\a\dot \a}$ with conjugate momenta $P_{\a\dot \a}$ and 4 from S$^5$ $Y^{a\dot a}, P_{a\dot a}$---and 8 complex fermions $\eta^{\a\dot a},\theta^{a\dot \a}$ which satisfy canonical (anti)commutation relations
\be
\begin{aligned}
[Y^{a\dot a}(\sigma,\tau),P_{b\dot b}(\sigma',\tau)]=i\, \delta^a_b\delta^{\dot a}_{\dot b}\, \delta(\sigma-\sigma'),
\qquad
\{\theta^{a\dot \a}(\sigma,\tau),\bar\theta_{b\dot \b}(\sigma',\tau)]= \delta^a_b\delta^{\dot \a}_{\dot \b}\, \delta(\sigma-\sigma'),
\end{aligned}
\ee
and similarly for $Z^{\a\dot\a}$ and $\eta^{\a\dot a}$.
Undotted and dotted indices refer to the two copies of the $\psu(2|2)$ superalgebra.
The (quantised) {}{Noether} charges corresponding to the two copies of $\psu(2|2)$ may be derived from the supercoset construction~\cite{Arutyunov:2006ak,Arutyunov:2009ga}\footnote{Our expressions differ from those of~\cite{Arutyunov:2009ga} for the coefficient in front of the fermionic contribution to the $\su(2)$ charges; we find that it should be $1/2$ in order to satisfy the correct commutation relations~\eqref{eq:comm-rel-AdS5-BMN}. Our conventions to raise and lower indices are also different.}
\be
\begin{aligned}
\gen{L}_a^{\ b}&=\int d\sigma\left[ -\tfrac{i}{2} \left( P_{a\dot c}Y^{b\dot c}-P^{b\dot c}Y_{a\dot c} \right)
+\tfrac{1}{2}\left(\bar \theta_{a\dot\gamma}\theta^{b\dot\gamma}-\bar\theta^{b\dot\gamma}\theta_{a\dot\gamma}\right)\right],\\
\gen{R}_\a^{\ \b}&=\int d\sigma\left[ -\tfrac{i}{2} \left( P_{\a\dot \gamma}Z^{\b\dot \gamma}-P^{\b\dot \gamma}Z_{\a\dot \gamma} \right)
+\tfrac{1}{2}\left(\bar \eta_{\a\dot c}\theta^{\b\dot c}-\bar\eta^{\b\dot c}\eta_{\a\dot c}\right)\right],\\
\gen{Q}_\a^{\ a}&=\tfrac{1}{2}e^{-i\pi/4}\int d \sigma\Big[
i P^{a\dot c}\bar\eta_{\a\dot c}+2Y^{a\dot c}\bar\eta_{\a\dot c}-2Y^{a\dot c}\eta_{\a\dot c}'
+P_{\a\dot\gamma}\theta^{a\dot\gamma}+2i Z_{\a\dot\gamma}\theta^{a\dot{\gamma}}-2i Z_{\a\dot\gamma}\bar\theta^{'a\dot\gamma}
\Big],\\
\overline{\gen{Q}}_a^{\ \a}&=\tfrac{1}{2}e^{i\pi/4}\int d \sigma\Big[
-i P_{a\dot c}\eta^{\a\dot c}+2Y_{a\dot c}\eta^{\a\dot c}-2Y_{a\dot c}\bar\eta^{'\a\dot c}
+P^{\a\dot\gamma}\bar\theta_{a\dot\gamma}-2i Z^{\a\dot\gamma}\bar\theta_{a\dot{\gamma}}+2i Z^{\a\dot\gamma}\theta_{a\dot\gamma}'
\Big].\\
\end{aligned}
\ee
The charges written above correspond to one copy of $\psu(2|2)$.\footnote{The expressions for the charges of the second copy are easily obtained from the above ones, since it is enough to swap the role of dotted/undotted indices and the role of the fermions $\theta$ and $\eta$.} All the charges above plus the Hamiltonian may be identified with specific elements of $\psu(2,2|4)$, the full isometry of superstring on $AdS_5\times S^5$. In particular, they correspond to charges which do not have an explicit dependence on the worldsheet time $\tau$, and therefore commute with $\gen{H}$.

The Hamiltonian and the worldsheet momentum are written as local integrals 
$$
\gen{H} = \int d\sigma\mathscr{H}, \qquad \gen{P} = -\int d\sigma\mathscr{P}
$$ 
of densities
\be
\begin{aligned}
\mathscr{H}&=\tfrac{1}{4}P_{a\dot a}P^{a\dot a}+Y_{a\dot a}Y^{a\dot a}+Y_{a\dot a}'Y^{'a\dot a}
+\tfrac{1}{4}P_{\a\dot \a}P^{\a\dot \a}+Z_{\a\dot \a}Z^{\a\dot \a}+Z_{\a\dot \a}'Z^{'\a\dot \a}\\
&+\bar\eta_{\a\dot a}\eta^{\a\dot a}+\tfrac{1}{2}\eta^{\a\dot a}\eta_{\a\dot a}'-\tfrac{1}{2}\bar\eta^{\a\dot a}\bar\eta_{\a\dot a}'
+\bar\theta_{a\dot \a}\theta^{a\dot \a}+\tfrac{1}{2}\theta^{a\dot \a}\theta_{a\dot \a}'-\tfrac{1}{2}\bar\theta^{a\dot \a}\bar\theta_{a\dot \a}',\\
\\
\mathscr{P}&=P_{a\dot a}Y^{'a\dot a}+P_{\a\dot \a}Z^{'\a\dot \a}
+i\bar\eta_{\a\dot a}\eta^{'\a\dot a}+i\bar\theta_{a\dot \a}\theta^{'a\dot \a}.
\end{aligned}
\ee
They correspond to invariance of the action under $\tau$- and $\sigma$-translations.
In the semiclassical limit the gauge-fixed action is invariant also under the  Lorentz boost, which acts infinitesimally on the worldsheet coordinates as $\delta \sigma=\varepsilon \tau,\ \delta\tau = \varepsilon \sigma$. Its {}{Noether} charge takes the form of a local integral
\be
\gen{J}=\int d\sigma \left(\sigma\mathscr{H}+\tau \mathscr{P}\right).
\ee
Because of the explicit $\tau$-dependence $\gen{J}$ does not commute with $\gen{H}$. Using the canonical commutation relations above, a straightforward computation of the commutators  shows that we indeed reproduce all commutation relations~\eqref{eq:comm-rel-J-AdS5-BMN} valid in the semiclassical limit. Obviously, the crucial role in this computation is played by the explicit $\sigma$-dependence in the integral representation of $\gen{J}$; in fact, the correct terms appear when first integrating by parts to get rid of derivatives of the $\delta$-function and then acting with $\partial_\sigma$ on $\sigma$.\footnote{When checking each commutation relation we  also have to identify all total-derivative terms that need to be subtracted.} This computation also confirms that the boost acts within each copy of $\psu(2|2)$ without mixing them.

It would be interesting to repeat the computation in the hybrid expansion of~\cite{Arutyunov:2006ak} to see if one can reproduce the full dependence on the exponential of the momentum.

\subsection{Limit $g\to 0$}
Another interesting limit of the superalgebra is  $g\to 0$, which corresponds to the weak coupling limit in $\mathcal N=4$ SYM. If we do not rescale any generator and we just set $g=0$ we find that the only (anti)-commutation relations that survive are
\be\label{eq:comm-rel-g-0}
\begin{aligned}
&\{\gen{Q}_\alpha^{\ a},\overline{\gen{Q}}_b^{\ \beta}\}= \delta^a_b \gen{R}_\alpha^{\ \beta}+\delta_\alpha^\beta\gen{L}_b^{\ a}+\tfrac{1}{2}\delta^a_b\delta_\alpha^\beta \gen{H},\qquad\qquad
[\gen{J},\gen{P}]=i\, \gen{H},
\end{aligned}
\ee
together with the commutation relations involving $\su(2)$ generators, which we do not repeat.
The above is an $\su(2|2)$ superalgebra, which shares the central element with the Heisenberg algebra spanned by $\{\gen J,\gen P, \gen H\}$.
In the fundamental representation, the energy is just a number independent of the momentum $h_p=1+\mathcal O(g^2)$, and since the coefficients in~\eqref{eq:par-repr} expand as $a_p=\bar a_p=1+\mathcal O(g)$ and  $b_p=\bar b_p=\mathcal O(g)$ the action of the supercharges becomes simply
\be
\gen{Q}_\a^{\ a}\ket{\chi_b}=\delta^a_b\ket{\chi_\a},
\qquad
\qquad
\overline{\gen{Q}}_{a}^{\ \a}\ket{\chi_\b}= \delta^{\a}_{\b}\ket{\chi_a}.
\ee
If we implement the $g\to 0$ limit on the coproducts of section~\ref{sec:boost} we see that they do not trivialise, in particular for the boost $\Delta(\gen{J})=\Delta'(\gen{J})+\mathcal{T}_{\gen{H}\hat{\gen{B}}}+\mathcal{T}_{\alg{psu}(2|2)}+\mathcal{T}_1$ we find
\begin{eqnarray}
&&\Delta'(\gen{J})=\gen{J}\otimes 1+1\otimes \gen{J}
- \frac{\sin p_1+\sin p_2-\sin (p_1+p_2)}{\sin p_1-\sin p_2}\left(\gen{J}\otimes 1-1\otimes \gen{J}\right)+\mathcal O(g), \nonumber\\
&&\mathcal{T}_{\gen{H}\hat{\gen{B}}}=\frac{1}{4} \frac{1}{\sin^{-1} p_1-\sin^{-1} p_2}\left(1-\tan\frac{p}{2}\otimes \tan\frac{p}{2}\right) \left(\gen{1}\otimes \hat{\gen{b}}+ \hat{\gen{b}}\otimes \gen{1}\right)+\mathcal O(g),\nonumber \\ 
&&\mathcal{T}_{\alg{psu}(2|2)}=\frac{1}{2}\frac{\sin p_2+\sin p_1}{\sin p_2-\sin p_1}\Big(\gen{K}^{-\frac{1}{4}} \gen{Q}_\a^{\ a}\otimes \gen{K}^{-\frac{1}{4}} \overline{\gen{Q}}_a^{\ \a}
-\gen{K}^{\frac{1}{4}} \overline{\gen{Q}}_a^{\ \a}\otimes \gen{K}^{\frac{1}{4}} \gen{Q}_\a^{\ a} \nonumber\\
&&\qquad\qquad\qquad\qquad\qquad\qquad+\gen{L}_a^{\ b}\otimes\gen{L}_{ b}^{\ a}
-\gen{R}_\a^{\ \b}\otimes\gen{R}_{ \b}^{\ \a}\Big)+\mathcal O(g),\nonumber
\end{eqnarray}
where $ \hat{\gen{b}}=\frac{1}{2}\cot\frac{p}{2}\mathbf{1}_g$.
On the other hand, one could just take all the coproducts to be the trivial ones, and that would give a homomorphism for the above (anti)commutation relations.
In this limit the boost is just represented by $\gen J=i\, \gen 1\partial_p$ and {we expect it to be connected to the operator used in~\cite{Bargheer:2008jt,Bargheer:2009xy} to generate long-range interactions starting from nearest-neighbour spin-chains. This should in turn be related to Baxter's general construction of the corner transfer matrix (cf. the account of \cite{Thacker:1985gz}), as it amounts to differentiation with respect to the uniformising rapidities associated to the lattice Boltzmann weigths.}

\vspace{12pt}

An alternative $g\to 0$ limit is obtained by  first rescaling $\gen J\to g\, \gen J$ and $\gen H\to g\, \gen H$ 
\be
\begin{aligned}
&\{\gen{Q}_\alpha^{\ a},\overline{\gen{Q}}_b^{\ \beta}\}= \delta^a_b \gen{R}_\alpha^{\ \beta}+\delta_\alpha^\beta\gen{L}_b^{\ a},\\
& [\gen{J},\gen{P}]=i\, \gen{H},\qquad\qquad
&&[\gen{J},\gen{Q}_\a^{\ a}]=-\tfrac{i}{4}\left(\gen{K}^{\frac{1}{2}}+\gen{K}^{-\frac{1}{2}}\right) \epsilon_{\a\b}\epsilon^{ab}\overline{\gen{Q}}_b^{\ \b},\\
&[\gen{J},\gen{H}]=\tfrac{1}{2}\left(\gen{K}-\gen{K}^{-1}\right) ,\qquad
&&[\gen{J},\overline{\gen{Q}}_a^{\ \a}]=-\tfrac{i}{4}\left(\gen{K}^{\frac{1}{2}}+\gen{K}^{-\frac{1}{2}}\right) \epsilon_{ab}\epsilon^{\a\b}\gen{Q}_\b^{\ b}.
\end{aligned}
\ee
The above still contains a $q$-deformed Poincar\'e subalgebra; the boost acts also on the supercharges, which are part of a $\alg{psu}(2|2)$ subalgebra. Further resclaing $\gen P\to \varepsilon\, \gen P$ and $\gen H\to \varepsilon\, \gen H$ and then sending $\varepsilon\to 0$ one obtains the classical superalgebra
\be
\begin{aligned}
&\{\gen{Q}_\alpha^{\ a},\overline{\gen{Q}}_b^{\ \beta}\}= \delta^a_b \gen{R}_\alpha^{\ \beta}+\delta_\alpha^\beta\gen{L}_b^{\ a},\\
& [\gen{J},\gen{P}]=i\, \gen{H},\qquad\qquad
&&[\gen{J},\gen{Q}_\a^{\ a}]=-\tfrac{i}{2}\epsilon_{\a\b}\epsilon^{ab}\overline{\gen{Q}}_b^{\ \b},\\
&[\gen{J},\gen{H}]=i\gen P,\qquad
&&[\gen{J},\overline{\gen{Q}}_a^{\ \a}]=-\tfrac{i}{2} \epsilon_{ab}\epsilon^{\a\b}\gen{Q}_\b^{\ b}.
\end{aligned}
\ee

\section{Cobracket}\label{sec:cobracket}
In this section we wish to study the first correction to the semiclassical result of the coproduct for the boost\footnote{We thank Niklas Beisert for a discussion which originated this idea, cf. also \cite{Beisert:2010kk}.}.
Let us remind that in order to implement the semiclassical limit in the fundamental representation it is convenient to parameterise
\be
x^\pm =x \left(\sqrt{1-\frac{x^2}{g^2
   \left(x^2-1\right)^2}}\pm\frac{i
   x}{g(1- x^2)}\right),
\ee
and then send $g\to\infty$. It will be also convenient to introduce the semiclassical spectral parameter $u$ which is related to $x$ as $u=x+1/x,\ x=\frac{1}{2}\left(u+\sqrt{u^2-4}\right)$. 

The $R$-matrix expands in the semiclassical limit as
\be
R=1+g^{-1}(r+r_0)+\mathcal{O}(g^{-2}),\qquad
r_0=\phi_0 \mathbf{1}\otimes \mathbf{1}.
\ee
In the following we will ignore $r_0$, which is proportional to the identity and sensitive to the normalisation for $R$. Similarly, we assume that at leading order in a semiclassical expansion the coproduct for the boost is the trivial one
\be
\Delta(\gen{J})=\gen{J}\otimes 1+1\otimes \gen{J}+g^{-1}\Delta_{(1)}(\gen{J})+\mathcal{O}(g^{-2}).
\ee
It can be checked that this is indeed the case in the fundamental representation, see~\eqref{eq:Delta-triv}. Then the fact that the boost is a symmetry of $R$ implies
\be
0=\Delta^{op}(\gen{J})R-R\Delta(\gen{J})
=g^{-1}\Big([\gen{J}\otimes 1+1\otimes \gen{J},r]-\delta(\gen{J})\Big)+\mathcal{O}(g^{-2})
\ee
where we defined $\delta(\gen{J})\equiv\Delta_{(1)}(\gen{J})-\Delta_{(1)}^{op}(\gen{J})$, which is the first non-trivial correction to $\Delta(\gen{J})$ that is antisymmetric under the action of $^{op}$. 
It is easy to check that the contribution $\Delta'(\gen{J})$ in~\eqref{eq:DeltaJ} is {}{symmetric} under the action of $^{op}$, and for this reason it is not captured by this computation.
The above equation states that $\delta(\gen{J})$ can be derived just by computing $[\gen{J}\otimes 1+1\otimes \gen{J},r]$; since $r$ satisfies the classical YBE, $\delta(\gen{J})$ is a cobracket and one obtains a quasi-triangular Lie bialgebra, see also~\cite{Beisert:2007ty}.

For the computation we will need the commutation relations between the boost and the other generators, see~\eqref{eq:comm-rel-J-AdS5-BMN} for those at level 0. However, here we will need also the action of $\gen{J}$ on  the generators at level $n$; if we write $\tilde{\gen{q}}_0\equiv[\gen{J},\gen{q}_0]$ where the subscript indicates the level, then we will take in the semiclassical limit
\be\label{eq:Jqn}
[\gen{J},\gen{q}_n]=\tilde{\gen{q}}_n+in\left(2\gen{q}_{n-1}-\tfrac{1}{2}\gen{q}_{n+1}\right),
\ee
which  is true at least in the evaluation representation.\footnote{In the evaluation representation we assume that $\gen{q}_n=u^n\gen{q}_0$ and $\gen{J}=ih_p \frac{du}{dp}\, \frac{d}{du} =\frac{i}{2}(4-u^2)\frac{d}{du}$. The boost will now act not just on the generator itself but also on the spectral parameter that multiplies it.} Additionally, we will assume that in the semiclassical limit the commutator with the secret symmetry\footnote{This commutation relation is inferred from the fundamental representation and we assume that it is universal in the semiclassical limit. Here $\gen{B}_1=\hat{\gen{B}}$, and at level 1 one would have $[\gen{J},\gen{B}_1]=-\frac{i}{2}\gen{B}_{2}$.} at level 0 is
\be
[\gen{J},\gen{B}_0]=-2i\, \gen{B}_{-1}.
\ee
We now use the universal r-matrix proposed in~\cite{Beisert:2007ty}, which in our conventions reads as
\be
r=-i\left(2\, r_{\gen{Q}\overline{\gen{Q}}}+2\, r_{\alg{su}(2)^2}+\, r_{\gen{H}\gen{B}}\right),
\ee
where 
\be\label{eq:r-univ}
\begin{aligned}
r_{\gen{Q}\overline{\gen{Q}}}=&
\sum_{m=0}^\infty \left[ (\gen{Q}_{-1-m})_\alpha^{\ a} \otimes (\overline{\gen{Q}}_{m})_a^{\ \alpha}
- (\overline{\gen{Q}}_{-1-m})_a^{\ \alpha}\otimes (\gen{Q}_{m})_\alpha^{\ a}\right], \\
r_{\alg{su}(2)^2}=&\sum_{m=0}^\infty \left[ (\gen{L}_{-1-m})_a^{\ b} \otimes (\gen{L}_{m})_b^{\ a}
- (\gen{R}_{-1-m})_\alpha^{\ \beta} \otimes (\gen{R}_{m})_\beta^{\ \alpha}\right],\\
r_{\gen{H}\gen{B}}=&
\sum_{m=-1}^\infty \gen{B}_{-1-m}\otimes \gen{H}_m +\sum_{m=1}^\infty \gen{H}_{-1-m}\otimes \gen{B}_m.
\end{aligned}
\ee
From~\eqref{eq:Jqn} we see that the action of the boost produces generators that either remain at the same level or  move by $\pm 1$ level. The contributions of the generators remaining at the same level will all cancel out in the computation for $r_{\gen{Q}\overline{\gen{Q}}}$. For $r_{\alg{su}(2)^2}$ the same holds trivially, essentially because the boost does not act on $\su(2)$ generators at level 0.
The result for the cobracket can be written as
\be
\begin{aligned}
\delta(\gen{J})=&-i\left(2\, \delta_{\gen{Q}\overline{\gen{Q}}}(\gen{J})+2\, \delta_{\alg{su}(2)^2}(\gen{J})+\, \delta_{\gen{H}\gen{B}}(\gen{J})\right),\\
\delta_{\gen{Q}\overline{\gen{Q}}}(\gen{J})=&
i\sum_{m=0}^\infty \left[ (\gen{Q}_{-m})_\alpha^{\ a} \otimes (\overline{\gen{Q}}_{m})_a^{\ \alpha}
- (\overline{\gen{Q}}_{-m})_a^{\ \alpha}\otimes (\gen{Q}_{m})_\alpha^{\ a}\right]\\
&-\frac{i}{2}\left[ (\gen{Q}_{0})_\alpha^{\ a} \otimes (\overline{\gen{Q}}_{0})_a^{\ \alpha}
- (\overline{\gen{Q}}_{0})_a^{\ \alpha}\otimes (\gen{Q}_{0})_\alpha^{\ a}\right],\\
\delta_{\alg{su}(2)^2}(\gen{J})=&
i\sum_{m=0}^\infty \left[ (\gen{L}_{-m})_a^{\ b} \otimes (\gen{L}_{m})_b^{\ a}
- (\gen{R}_{-m})_\alpha^{\ \beta} \otimes (\gen{R}_{m})_\beta^{\ \alpha}\right]\\
&-\frac{i}{2} \left[ (\gen{L}_{0})_a^{\ b} \otimes (\gen{L}_{0})_b^{\ a}
- (\gen{R}_{0})_\alpha^{\ \beta} \otimes (\gen{R}_{0})_\beta^{\ \alpha}\right],\\
\delta_{\gen{H}\gen{B}}(\gen{J})=&i\Bigg(
-2\sum_{m=-2}^\infty \gen{B}_{-2-m}\otimes \gen{H}_m +\sum_{m=0}^\infty \gen{B}_{-m}\otimes \gen{H}_m 
+\sum_{m=-1}^\infty \gen{B}_{-1-m}\otimes \gen{P}_m\\
&-2\sum_{m=1}^\infty \gen{H}_{-2-m}\otimes \gen{B}_m+\sum_{m=1}^\infty \gen{H}_{-m}\otimes \gen{B}_m
+\sum_{m=1}^\infty \gen{P}_{-1-m}\otimes \gen{B}_m
\Bigg).
\end{aligned}
\ee
The computation of $\delta_{\gen{H}\gen{B}}(\gen{J})$ crucially depends on the fact that the sums in~\eqref{eq:r-univ} start from $m=-1$ and $m=1$ rather than $m=0$.
We also notice that in the evaluation representation we can identify $\gen{P}_n \sim 2\gen{H}_{n-1}$, so that 
\be
\begin{aligned}
\delta_{\gen{H}\gen{B}}(\gen{J}) \sim &
i\Bigg(\sum_{m=0}^\infty \gen{B}_{-m}\otimes \gen{H}_m
+\sum_{m=1}^\infty \gen{H}_{-m}\otimes \gen{B}_m\Bigg).
\end{aligned}
\ee
It is interesting to write the  result for the cobracket in the evaluation representation. We obtain
\be
\begin{aligned}
\delta(\gen{J})&=
\frac{u_1+u_2}{u_1-u_2}\Bigg[ \gen{Q}_\alpha^{\ a} \otimes \overline{\gen{Q}}_a^{\ \alpha}
- \overline{\gen{Q}}_a^{\ \alpha}\otimes \gen{Q}_\alpha^{\ a}
+ \gen{L}_a^{\ b} \otimes \gen{L}_b^{\ a}
- \gen{R}_\alpha^{\ \beta} \otimes \gen{R}_\beta^{\ \alpha}\Bigg]\\
&+\frac{1}{u_1-u_2}\left( \gen{B}_{1}\otimes \gen{H}
+ \gen{H}\otimes \gen{B}_1\right).
\end{aligned}
\ee
Surprisingly, this is very suggestive of the form of the coproduct in~\eqref{eq:DeltaJ}. Together with other modifications, it seems that the role of the semiclassical parameter $u$ is played at all loops {}{by} $w$. Notice that both $w=2(1+x^+x^-)/(x^++x^-)$ and $\hat u=(x^++1/x^++x^-+1/x^-)/2$ reduce in the semiclassical limit to $u=x+1/x$.

{Let us connect our results to those of \cite{Beisert:2010kk}. In that paper, the classical limit of the {\it quantum deformed} $R$-matrix of \cite{Beisert:2008tw} was analysed, and a classical {\it trigonometric} $r$-matrix was obtained. In that situation, the semiclassical limit singles out a trigonometric spectral variable $z$, which allows to re-express the algebra in a series of affine levels, in the spirit of Drinfeld's second realisation of quantum algebras. A universal form of the classical $r$-matrix was achieved, and a {\it derivation}
\begin{equation}
D = z \frac{d}{dz}
\end{equation}
was introduced, traditionally counting the levels. However, the constraints of the problem force $z$ to depend on the remaining representation labels, making $D$ no longer a mere level-counter, but rather equipped with non-trivial commutation rules and cobracket. Beisert then reduced the construction to the rational case by taking a further limit on the parameters, obtaining a rescaled generator he called $\tilde{D}$, acting as a derivative with respect to the classical rational spectral variable $u$
\begin{equation}
\tilde{D} = \frac{d}{du}.
\end{equation} 
He observed that this generator should  naturally have non-trivial cobracket, and commented upon the difficulty of rendering this a symmetry of the quantum $S$-matrix.} 

{We  notice\footnote{We thank Niklas Beisert for important discussions about this point.} that, when taken in the classical limit, our boost generator acts in a very similar fashion to $\tilde{D}$, differing only by a $u$-dependent prefactor related to the particle energy. We expect that the cobracket of $\tilde{D}$ should structurally be very similar to the one we find for the boost. Since only a rescaling discriminates between them, we argue that our quantum boost symmetry could morally be considered as a quantisation of Beisert's derivation.}

\section{A contraction of $U_q(\alg{d}(2,1;\alpha))$}\label{sec:q-d21a}
In Section~\ref{sec:semcl} we reviewed the contraction that reproduces the classical Poincar\'e superalgebra under study from $\alg{d}(2,1;\varepsilon)$. It is natural to ask whether a contraction of $U_q(\alg{d}(2,1;\varepsilon))$ \cite{Beisert:2017xqx,zou1996deformation,Matsumoto:2008ww,Heckenberger:2007ry} can reproduce the $q$-deformed Poincar\'e superalgebra. This is an interesting question since it may open the possibility of looking at the contraction also in the affine case, and obtain a quantum affine Poincar\'e superalgebra that extends the one of Section~\ref{sec:q-affine}.

To construct $U_q(\alg{d}(2,1;\varepsilon))$ we first choose a Serre-Chevalley basis for the undeformed superalgebra. We prefer to work in the grading where all simple generators are fermionic
\be\label{eq:simpl}
\begin{aligned}
&\gen h_1 ={\phantom{-}} (1-\varepsilon)\gen{L}_2^{\ 2}+\gen{R}_4^{\ 4}-\varepsilon \gen{J}_6^{\ 6},\qquad
&&\gen e_1=\gen{S}^{146},\qquad
&&&\gen f_1=\gen{S}^{235},\\
&\gen h_2 = -(1-\varepsilon)\gen{L}_2^{\ 2}-\gen{R}_4^{\ 4}-\varepsilon \gen{J}_6^{\ 6},\qquad
&&\gen e_2=\gen{S}^{236},\qquad
&&&\gen f_2=\gen{S}^{145},\\
&\gen h_3 = -(1-\varepsilon)\gen{L}_2^{\ 2}+\gen{R}_4^{\ 4}+\varepsilon \gen{J}_6^{\ 6},\qquad
&&\gen e_3=\gen{S}^{245},\qquad
&&&\gen f_3=\gen{S}^{136}.\\
\end{aligned}
\ee
They satisfy
\be\label{eq:comm-rel-sc}
[\gen h_i,\gen h_j]=0,\qquad
[\gen h_i,\gen e_j]=a_{ij}\gen e_j,\qquad
[\gen h_i,\gen f_j]=-a_{ij}\gen f_j,\qquad
\{\gen e_i,\gen f_j\}=\delta_{ij}\gen h_i,
\ee
where $i,j=1,2,3$ and the Cartan matrix is
\be
a_{ij}=\left(
\begin{array}{ccc}
 0 & \varepsilon  & -1 \\
 \varepsilon  & 0 & 1-\varepsilon  \\
 -1 & 1-\varepsilon  & 0 \\
\end{array}
\right).
\ee
Notice that we have fixed the simple generators in order to have a {}{symmetric} Cartan matrix. The remaining generators of $\alg{d}(2,1,\varepsilon)$  may be obtained by taking {}{appropriate} (anti)commutators of the simple ones.
In particular, to recover the remaining bosonic generators we define $\gen e_{ij}\equiv a_{ij}^{-1}\{e_i,e_j\}$, $\gen f_{ij}\equiv -a_{ij}^{-1}\{f_j,f_i\}$ and we take
\be\label{eq:non-simpl-bos}
\begin{aligned}
&\gen{J}_5^{\ 6 }=\gen e_{12},\qquad
&&\gen{R}_3^{\ 4 }=\gen e_{13},\qquad
&&&\gen{L}_1^{\ 2 }=\gen e_{23},
\\
&\gen{J}_6^{\ 5 }=\gen f_{12},\qquad
&&\gen{R}_4^{\ 3 }=\gen f_{13},\qquad
&&&\gen{L}_2^{\ 1 }=\gen f_{23},
\end{aligned}
\ee
while the two remaining fermionic generators are written in terms of $\gen e_{123}\equiv[\gen e_{12},\gen e_3]$, $\gen f_{123}\equiv-[\gen f_3,\gen f_{12}]$
\be\label{eq:non-simpl-fer}
\gen{S}^{246}=-\gen e_{123},\qquad\quad
\gen{S}^{135}=-\gen f_{123}.
\ee
If we use the above identifications, together with $\gen e_i^2=0=\gen f_i^2, i=1,2,3$ and the relations in~\eqref{eq:comm-rel-sc} we reproduce the commutation relations of $\alg{d}(2,1;\varepsilon)$.

\vspace{12pt}

Let us now quantise the superalgebra. This is achieved by first modifying the (anti)commutation relations for simple generators as
\be\label{eq:q-comm-rel-sc}
[\gen h_i,\gen h_j]=0,\qquad
[\gen h_i,\gen e_j]=a_{ij}\gen e_j,\qquad
[\gen h_i,\gen f_j]=-a_{ij}\gen f_j,\qquad
\{\gen e_i,\gen f_j\}=\delta_{ij}\frac{\gen k_i-\gen k_i^{-1}}{q-q^{-1}},
\ee
where we defined~\footnote{Notice that the factors of $q$, appearing when taking the anticommutator of a positive and a negative simple root,  are not raised to different exponents because we chose to have a {}{symmetric} Cartan matrix.} $q=e^{\frac{\hbar}{2}}$ and $\gen{k}_i=\exp(\frac{\hbar}{2}\gen h_i)$. 
Notice that $[\gen h_i,\gen q]=\alpha \gen q$ implies $\gen k_i\gen q=q^\alpha \gen q\gen k_i$.
Sending the deformation parameter $\hbar\to 0$ one recovers the $\alg{d}(2,1;\varepsilon)$ superalgebra. 

In analogy with the undeformed case, the remaining generators are obtained by taking {}{appropriate} combinations of the simple generators. We will now need to introduce the graded $q$-commutator $\dsbl,\dsbr_c$
\be
\dsbl A,B\dsbr_c= AB-(-1)^{|A|\cdot|B|}\, c\, BA,
\ee
where the degree is $|A|=0$ for bosonic and $|A|=1$ for fermionic generators.\footnote{When we do not write a subscript we mean  $c=1$, i.e. the usual (graded) commutator.}
We use it to define\footnote{These considerations are generic, and are valid for a quantum group with defining relations~\eqref{eq:q-comm-rel-sc}. In other words, we are not using the particular expression for the Cartan matrix, although we are obviously assuming that $a_{ij}\neq 0$.}
\be
\begin{aligned}
&\gen e_{ij}\equiv {\phantom{-}}[a_{ij}]_q^{-1}\dsbl \gen e_i,\gen e_j\dsbr_{q_{ij}}, \qquad\quad 
&&\gen e_{ijk}\equiv {\phantom{-}} \dsbl \gen e_{ij},\gen e_k\dsbr_{q_{ik}q_{jk}},\\
&\gen f_{ij}\equiv -[a_{ij}]_q^{-1}\dsbl \gen f_j,\gen f_i\dsbr_{q_{ij}^{-1}},
&&\gen f_{ijk}\equiv -\dsbl \gen f_k, \gen f_{ij}\dsbr_{q_{ik}^{-1}q_{jk}^{-1}},
\end{aligned}
\ee 
where we defined the $q$-number $[m]_q=\frac{q^m-q^{-m}}{q-q^{-1}}$ and the short-hand notation $q_{ij}=q^{a_{ij}}$.
These elements obviously reduce to the definitions used in the undeformed case in the $\hbar \to 0$ limit.
The reason to introduce the above elements is that they satisfy
\be
\begin{aligned}
&\dsbl \gen e_{ij},\gen f_{ij}\dsbr
=\frac{\gen k_{ij}-\gen k_{ij}^{-1}}{q_{ij}-q_{ij}^{-1}},\qquad
&&\dsbl \gen e_{ijk},\gen f_{ijk}\dsbr=\frac{[a_{ik}+a_{jk}]_q}{[a_{ij}]_q}\frac{\gen k_{ijk}-\gen k_{ijk}^{-1}}{q-q^{-1}}\\
&[\gen h_{k},\gen e_{ij}]={\phantom{-}}(a_{ki}+a_{kj})\gen e_{ij},\qquad
&&[\gen h_l,\gen e_{ijk}]={\phantom{-}}(a_{li}+a_{lj}+a_{lk})\gen e_{ijk},\\
&[\gen h_{k},\gen f_{ij}]=-(a_{ki}+a_{kj})\gen f_{ij},\qquad
&&[\gen h_l,\gen f_{ijk}]=-(a_{li}+a_{lj}+a_{lk})\gen f_{ijk}.
\end{aligned}
\ee
where we used the notation $\gen k_{i_1\cdots i_n}\equiv \gen k_{i_1}\cdots \gen k_{i_n}$. 
The $q$-deformed superalgebra $U_q(\alg{d}(2,1;\varepsilon))$ is then spanned by the simple generators $\gen e_i,\gen f_i$, $i=1,2,3$, the Cartan $\gen h_i$, and the non-simple generators $\gen e_{ij}, \gen f_{ij}$ with $ij=\{12,13,23\}$ and $\gen e_{123},\gen f_{123}$.

The deformed superalgebra contains three copies of $q$-deformed $\su(2)$. Each of them is generated by $\gen e_{ij},\gen f_{ij}, \gen k_{ij}$, and  comes with its own deformation parameter $q_{ij}$. Let us look at one such copy
\be
[\gen h_{12},\gen e_{12}]=2\varepsilon\, \gen e_{12},\qquad
[\gen h_{12},\gen f_{12}]=-2\varepsilon\, \gen f_{12},\qquad
[ \gen e_{12},\gen f_{12}]=\frac{\gen k_{12}-\gen k_{12}^{-1}}{q^\varepsilon-q^{-\varepsilon}}.
\ee
If we identify the generators\footnote{This identification coincides with the one in the undeformed case (cf.~\eqref{eq:cl-d21a-sP},~\eqref{eq:simpl} and~\eqref{eq:non-simpl-bos}) if we identify the deformation parameter $\hbar=2/g$. The extra powers of $g$ are due to the rescalings~\eqref{eq:rescale-JP} implemented before taking the semiclassical limit.} as
\be\label{eq:ex-id-qP}
\gen{H}=-\frac{i}{2}\varepsilon\hbar g\left(\gen e_{12}-\gen f_{12}\right),\qquad
\gen{J}=-\frac{i}{2}g\left(\gen e_{12}+\gen f_{12}\right),\qquad
\gen{P}=-\frac{i}{2}\hbar \gen h_{12},
\ee  
so that $\gen K=\gen k_{12}$, and then send $\varepsilon\to 0$ we recover the commutation relations of the bosonic $q$-Poincar\'e algebra as in~\eqref{eq:comm-rel-J-AdS5}.
 Notice that here we did not need to take the semiclassical $\hbar\to 0$ limit. We expect that we will eventually have to send $\hbar \to 0$ in order to remove the deformation in the other two copies of $q$-deformed $\su(2)$; in fact after sending $\varepsilon\to 0$ these are still deformed, with parameters $q$ and $q^{-1}$. On the other hand, the contraction $\hbar\to  0$ can be safely taken on the bosonic $q$-Poincar\'e  subalgebra, because the explicit $\hbar$ does not appear in its commutation relations.

\vspace{12pt}

Let us now try to include also the supercharges in the discussion. Here we will just sketch a calculation to highlight an apparent obstruction in obtaining the $q$-Poincar\'e superalgebra as a contraction of $U_q(\alg{d}(2,1;\varepsilon))$. Let us for example consider the supercharges $\gen Q^{41},\gen Q^{32}$ and $\overline{\gen Q}^{23}$; in the semiclassical limit they are just written as linear combinations of the simple generators $\gen e_i, \gen f_i,\ i=1,2$, see also~\eqref{eq:cl-d21a-sP}
\be
\begin{aligned}
\gen Q^{41}=\frac{1}{\sqrt 2}\left(\gen f_2-i\gen e_1\right),\qquad
\gen Q^{32}=\frac{1}{\sqrt 2}\left(\gen f_1-i\gen e_2\right),\qquad
\overline{\gen Q}^{23}=\frac{1}{i\sqrt 2}\left(\gen f_1+i\gen e_2\right).
\end{aligned}
\ee
We now write their anticommutators (in the $\varepsilon\to 0$ limit) and focus for simplicity just on the contribution of the Cartan generators $\gen h_i$
\be
\begin{aligned}
\{\gen Q^{41},\gen Q^{32}\}&\sim \gen h_{1}+\gen h_{2}+\cdots\to \gen P+\cdots,\\
\{\gen Q^{41},\overline{\gen Q}^{23}\}&\sim \gen h_{1}-\gen h_{2}+\cdots\to \gen L_2^{\ 2}+\gen R_4^{\ 4}+\cdots
\end{aligned}
\ee
In the deformed case the contribution of the Cartan generators is traded for expressions in terms of $\gen k_i$. This obviously matches with the fact that we want the anticommutator $\{\gen Q^{41},\gen Q^{32}\}$ to give the {}{exponential} of the momentum, i.e.  we  want the appearance of (powers of) the generator $\gen k_{12}=\gen k_1 \gen k_2$. It is clear that the above identification of the supercharges will not work in the deformed case, since their anticommutator would give terms proportional to the {}{sum} of $\gen k_1$ and $\gen k_2$ (and their inverses) rather than their products as wanted. This problem is however easily solved by dressing the fermionic generators with {}{appropriate} powers of $\gen k_i$. We could take for example\footnote{The above is just one of the possible identifications that produce $\gen k_{12}-\gen k_{12}^{-1}$ when taking the anticommutator of the supercharges, and we choose this just to present an explicit example. Notice that since this choice leads to identify $\gen k_{12}=\gen K^{\frac{1}{2}}$, it is not consistent with the one in~\eqref{eq:ex-id-qP} for the $q$-Poincar\'e subalgebra, which should then be modified. This will not matter for our argument.}
\be
\begin{aligned}
\gen Q^{41}=\frac{1}{\sqrt 2}\left(\gen f_2\gen k_1^{-\frac{1}{2}}-i\gen k_2^{\frac{1}{2}}\gen e_1\right),
\gen Q^{32}=\frac{1}{\sqrt 2}\left(\gen f_1\gen k_2^{\frac{1}{2}}-i\gen k_1^{-\frac{1}{2}}\gen e_2\right),
\overline{\gen Q}^{23}=\frac{1}{i\sqrt 2}\left(\gen f_1\gen k_2^{\frac{1}{2}}+i\gen k_1^{\frac{3}{2}}\gen e_2\right),
\end{aligned}
\ee
so that
\be
\begin{aligned}
\{\gen Q^{41},\gen Q^{32}\}&=-\frac{i}{2}\frac{\gen k_{12}-\gen k_{12}^{-1}}{q-q^{-1}}+\text{rest},\\
\{\gen Q^{41},\overline{\gen Q}^{23}\}&=-\frac{1}{2}\frac{\gen k_{1|2}-\gen k_{1|2}^{-1}}{q-q^{-1}}+\text{rest}.
\end{aligned}
\ee
After taking the $\varepsilon\to 0$ limit we still have that $\gen k_{12}$ and $\gen k_{1|2}=\gen k_1\gen k_2^{-1}$ are {}{exponentials} of $\gen P$ and $\gen L_2^{\ 2}+\gen R_4^{\ 4}$ respectively. Although this is fine for the momentum generator, we know that the $q$-Poincar\'e superalgebra contains two copies of classical $\su(2)_{\gen L}$, $\su(2)_{\gen R}$, meaning that we want to implement the semiclassical limit $\hbar \to 0$ in the second anticommutator. This would however create problems in the first anticommutator; in fact, since we identify $\gen k_{12}=\gen K^{\frac{1}{2}}$ in order to match with~\eqref{eq:comm-rel-noJ-AdS5}, the limit would produce a divergence coming from the denominator $q-q^{-1}$. On the other hand,  trying to remove this divergence by rescaling the supercharges with {}{appropriate} powers of $\hbar$ also does not work, since it would spoil $\{\gen Q^{41},\overline{\gen Q}^{23}\}$.

It would be nice to see if techinques similar to the ones used in~\cite{Beisert:2017xqx} may be of help to overcome these divergences.

\section{$q$-affine Poincar\'e}\label{sec:q-affine}
In this section we focus on the bosonic Poincar\'e algebra and obtain a quantum affine version of it; we get it as a contraction of the quantum affine algebra $U_q(\widehat{\alg{sl}_2})$ (see also \cite{Doikou:2009js}). {This is in the spirit of exploring extensions of the boost symmetry to higher levels, hoping to eventually make contact with the Yangian picture.
Notice that this idea would be in contrast with the interpretation of our boost as the quantisation of the derivation  appeared in \cite{Beisert:2010kk}, for which there are no higher partners.}

We start with the defining relations of $U_q(\widehat{\alg{sl}_2})$ in Drinfeld's second realisation (reported for convenience in \cite{deLeeuw:2011fr}, see also references therein). The generators in this realisation are $\gen{h}_n$, $\gen{e}^\pm_n$, $n \in \mathbb{Z}$, subject to 
\begin{eqnarray}\label{eq:def-rel-Uqsl2}
&&[\gen{h}_m,\gen{h}_n]=0, \nonumber \\
&&[\gen{h}_0, \gen{e}^\pm_n]=\pm 2 \, \gen{e}^\pm_{n}, \qquad [\gen{h}_m, \gen{e}^\pm_n]=\pm \frac{[2 m]_q}{m} \, \gen{e}^\pm_{m+n}, \ m\neq 0 \nonumber \\
&&[\gen{e}^+_m, \gen{e}^-_n] = \frac{1}{q-q^{-1}} \, \Big(\bpsi^+_{m+n} - \bpsi^-_{m+n}\Big), \\ 
&&[\gen{e}^\pm_{m+1}, \gen{e}^\pm_n]_{q^{\pm 2}} = - [\gen{e}^\pm_{n+1}, \gen{e}^\pm_m]_{q^{\pm 2}},\nonumber 
\end{eqnarray}
where $[x]_q\equiv \frac{q^x-q^{-x}}{q-q^{-1}}$ and $[\gen{x},\gen{y}]_q\equiv \gen{x}\gen{y}-q\gen{y}\gen{x}$, and  one needs to extract the generators $\bpsi^\pm_m$ from the formula
\begin{eqnarray}
\bpsi^\pm (z) = q^{\pm \gen{h}_0} \, \exp \bigg[ \pm \big(q - q^{-1}\big) \sum_{m>0} \gen{h}_{\pm m} \, z^{\pm m}\bigg] \, = \, \sum_{m \in \mathbb{Z}} \bpsi^\pm_m \, z^{m}.
\end{eqnarray}
Inspired by the finite case~\cite{Celeghini:1990bf} we define
\begin{eqnarray}
\gen{H}_m = \mu\varepsilon (\gen{e}^+_m + \gen{e}^-_m), \quad \gen J_m=\tfrac{1}{2}(\gen{e}^+_m - \gen{e}^-_m)  , \quad \gen P_m=-i\mu\varepsilon\, \gen{h}_m , \quad q = e^{\varepsilon \, \mu}, \quad \varepsilon \to 0.
\end{eqnarray}
In principle one may study different contractions where the powers of $\varepsilon$ depend also on the level $n$.
We refer to Appendix~\ref{app:q-aff} for the details on how to take this contraction on the quantum affine algebra. In fact, we find that implementing na\"ively the $\varepsilon\to 0$ limit on the $q$-Serre relations ultimately leads to commutation relations which are not consistent with the Jacobi identity. We therefore first solve the $q$-Serre relations and only later send $\varepsilon\to 0$, and we obtain
\begin{subequations}\label{eq:comm-rel-q-aff-poin}
\begin{align}
&[\gen P_m,\gen P_n]=0, \qquad [\gen P_m, \gen H_n]=0, \qquad [\gen H_m,\gen H_n]=0,  \label{eq:comm-rel-q-aff-poin-a} \\ 
&[\gen J_m, \gen P_n] = i\, \gen H_{m+n}, \label{eq:comm-rel-q-aff-poin-b} \\
&[\gen J_m,\gen H_n]=\frac{1}{2}\left(\bpsi^+_{m+n}-\bpsi^-_{m+n}\right)+\frac{1}{2}\, \text{sign}(m-n)\left(\sum_{\ell=0}^{m-n}\gen H_{m-\ell}\gen H_{n+\ell}-\gen H_m\gen H_n\right),  \label{eq:comm-rel-q-aff-poin-c} \\
&[\gen J_m,\gen J_n]= \frac{1}{4}\, \text{sign}(m-n)\Bigg(\sum_{\ell=0}^{m-n}\left(\gen H_{m-\ell}\gen J_{n+\ell}+\gen J_{m-\ell}\gen H_{n+\ell}
+\gen H_{n+\ell}\gen J_{m-\ell}+\gen J_{n+\ell}\gen H_{m-\ell}\right) \nonumber\\
&\qquad\qquad\qquad\qquad-(\gen H_n\gen J_m+\gen J_n\gen H_m+\gen H_m\gen J_n+\gen J_m\gen H_n)\Bigg).  \label{eq:comm-rel-q-aff-poin-d} 
\end{align}
\end{subequations}
When writing the sum $\sum_{\ell=0}^{m-n}$ we use the convention that $\ell=0,1,\ldots, m-n$ for $m>n$, while we let it run in the opposite direction $\ell=m-n,m-n+1,\ldots,-1,0$ for $m<n$.
Moreover, the generators $\bpsi^\pm_m$ should be extracted from
\begin{eqnarray}
\bpsi^\pm (z) = e^{\pm i \, \gen P_0} \, \exp \bigg[ \pm 2 i \sum_{m>0} \gen P_{\pm m} \, z^{\pm m}\bigg] \, = \, \sum_{m \in \mathbb{Z}} \bpsi^\pm_m \, z^{m}.
\end{eqnarray}
This we take as the quantum affine version of the Poincar\'e algebra.
Notice that no explicit deformation parameter appears here, in agreement with the comment after~\eqref{eq:comm-rel-J-AdS5}. If we want to match with~\eqref{eq:comm-rel-J-AdS5} it is enough to rescale $\gen H_n\to \gen H_n/g, \gen J_n\to \gen J_n/g$, at least at level 0.

From the above commutation relations one easily finds $(\gen P_\pm\equiv \sum_{n>0}\gen P_{\pm n}z^{\pm n})$
\be
\begin{aligned}
&[\gen J_m,\gen P_n^N]=iN\gen H_{m+n}\, \gen P_n^{N-1},\qquad\qquad
&&[\gen J_m,e^{\pm 2i \gen P_\pm}] = \mp 2 \left(\sum_{n>0}z^{\pm n}\gen H_{m\pm n}\right)\, e^{\pm 2i \gen P_\pm},
\\
&[\gen J_m,e^{\alpha \gen P_n}]=i\alpha \gen H_{m+n}\, e^{\alpha \gen p_n},
&&[\gen J_m,\bpsi^\pm(z)]=\mp \left(\gen H_m+2\left(\sum_{n>0}z^{\pm n}\gen H_{m+n}\right)\right)\bpsi^\pm (z).
\end{aligned}
\ee
It is useful to notice that $\bpsi^+(z)$ expands in non-negative powers of $z$, while $\bpsi^-(z)$ non-positive. This means that only $\bpsi^\pm_{\pm m}$ with $m\geq 0$ are non-vanishing ($\bpsi^\pm(z)=\sum_{n\geq 0}\bpsi^\pm_{\pm n}z^{\pm n}$) and that
\be
[\gen J_m,\bpsi^\pm_{\pm n}]=\mp  \gen H_m\, \bpsi^\pm_{\pm n}\mp 2\sum_{\substack{0\leq p<n}}\gen H_{m\pm (n-p)}\, \bpsi^\pm_{\pm p}.
\ee
The semiclassical limit is achieved by first rescaling $\gen P_m\to \mu\gen P_m, \gen H_m\to \mu\gen H_m$ and then sending $\mu\to 0$. We get
\be
\bpsi^\pm_0\sim \pm i\mu \, \gen P_0+\mathcal{O}(\mu^2),\qquad\qquad
\bpsi^\pm_{\pm m} \sim \pm 2i\mu \, \gen P_m+\mathcal{O}(\mu^2),\qquad m> 0,
\ee
so that commutation relations are simply
\be
[\gen J_m, \gen P_n]  = i\,  \gen H_{m+n},\qquad\qquad
[\gen J_m, \gen H_n]  =  i\,  \gen P_{m+n},
\ee
for all $m,n$.

{}{As reported in \cite{ding2000factorization}, there are two possible coproducts one can endow the quantum affine $U_q(\widehat{\alg{sl}_2})$ algebra with, which we might attempt to contract to obtain a well-defined coalgebra structure on $q$-affine Poincar\'e. On the one hand, the standard coproduct (\cite{ding2000factorization}, formula 2.8) is prohibitively complicated to study in the contraction, and we are unable to ascertain whether it may produce a finite result. On the other hand, the so-called ``Drinfeld coproduct'' (formula (3.2) in \cite{ding2000factorization}) is much simpler, but it seems to display divergences when we take the linear combinations which are appropriate to our contraction procedure. This seems to be in consonance with the expectation that the universal $R$-matrix will develop singularities in the contraction as well. If a coproduct exists, we believe that it should be obtained directly working with the limiting algebra, but we have not yet been able to find one due to the complicated form of the commutation relations.}

{In order to make direct use of this construction for the $AdS_5$ boost we would need an extension to superalgebras, ({\it e.g} along the lines of \cite{Gade:1997ltl}, see also \cite{Gade:2002eas}). This proves challenging at the moment, and we plan to return to it in future work.}

\section{Conclusions}

{In this paper we have discovered a new exact quantum symmetry of the $AdS_5 \times S^5$ string $S$-matrix in the fundamental representation. In our setup we  fixed the boost coproduct by demanding that it gives a symmetry of the scattering processes, therefore the dressing phase features as an input and cannot be rediscovered from the knowledge of the boost, {even when requiring the compatibility of the boost with the antipode of Hopf algebras.} However, we expect that a universal formulation of the full quantum group symmetry would fix uniquely the tail of the boost coproduct, and would therefore  produce a way, alternative to crossing symmetry, to determine the dressing phase.}
{The study of the boost symmetry in bound state representations may turn out to be helpful in the identification of the universal formulation.}

{It is crucial to remark that\footnote{{We thank Niklas Beisert for important discussions about this point.}} in the paper \cite{Beisert:2010kk} Beisert studied the classical limit of the {\it quantum deformed} $R$-matrix \cite{Beisert:2008tw}, obtaining a classical {\it trigonometric} $r$-matrix. This, we remind, entails a superimposed deformation with respect to our situation. Beisert introduced a {\it derivation}, originating from the quantum affine algebra underlying the quantum deformed $R$-matrix, and considered its classical limit. He then reduced it to the rational case by taking a further limit on the parameters, hence obtaining a generator dubbed $\tilde{D}$, acting as a derivative with respect to the classical spectral variable $u$. He anticipated that this generator should have non-trivial cobracket, and commented upon the ensuing difficulty of promoting it to a symmetry of the quantum problem. Our boost generator acts in a very similar fashion to $\tilde{D}$ in the classical limit, differing only by a $u$-dependent factor before the derivative. We expect the cobracket of $\tilde{D}$, which was not explicitly given in \cite{Beisert:2010kk}, to be structurally analogous to the one we find for the boost. We then believe that our quantum boost symmetry should morally be regarded as a quantisation of Beisert's derivation, should this be promoted to an exact quantum symmetry. 
Notice however that in \cite{Beisert:2010kk} $\tilde{D}$ is not part of the quantum affine algebra, it rather acts as an external generator. This would be in constrast with an alternative possibility pursued in section~\ref{sec:q-affine}, where the boost generator has higher level counterparts and is part of the quantum affine algebra.}

{On the spin-chain side, boost operators  have been altogether intensively studied \cite{tetelman1982lorentz,sogo1983boost,Bargheer:2008jt}, see also \cite{Beisert:2007jv,Zwiebel:2008gr,Beisert:2016qei,Klose:2016uur,Klose:2016qfv,Beisert:2017xqx}.  It would be very exciting to find new relationships with these approaches.}

Very recently, deformed Poincar\'e symmetries have featured  in the work of \cite{Beisert:2017xqx}, where a  novel deformation of 3D (rather than 2D) {\it kappa-Poincar\'e} was found, relying on the superimposed quantum deformation of the scattering problem already mentioned. 
{It would be very interesting to understand whether it is possible to combine the methods of \cite{Beisert:2017xqx} with the contraction of $\alg{d}(2,1;\alpha)$ used in our paper and in~\cite{Young:2007wd}---which produces a Poincar\'e superalgebra rather than a triple central extention of $\alg{psu}(2|2)$---to bridge the two approaches and overcome the issues of section~\ref{sec:q-d21a}.}

{In many respects, the way the boost symmetry manifests itself reminds of how the secret (or bonus) symmetry was first investigated. Both invariances are very well hidden, and it is both mathematical consistency and a search for a complete algebraic picture of the scattering problem which calls for them, and eventually leads {their discovery}. The bonus symmetry has by now been investigated in a variety of contexts; it was observed in boundary scattering problems \cite{Regelskis:2011fa}, $n$-point amplitudes \cite{Beisert:2011pn}, the pure-spinor formalism \cite{Berkovits:2011kn}, in the quantum-affine deformations \cite{deLeeuw:2011fr} and in the context of Wilson loops \cite{Munkler:2015xqa}. This makes it quite a significant feature of the system and not an isolated instance \cite{deLeeuw:2012jf}. It would be interesting to explore similar investigations also for the boost.}

{Finally,  non-standard quantum algebras and associated bonus generators analogous to those of $AdS_5/CFT_4$ have been found in lower-dimensional $AdS/CFT$ as well. All these settings share peculiar algebraic features stemming from the vanishing of the Killing form of their superisometry \cite{Zarembo:2010sg,Wulff:2015mwa}, dictated by string coset integrability {and 1-loop scale invariance}.  
From a quantum-group viewpoint, the integrable structure behind the $AdS_4$ case is reduced for the most part to the five-dimensional case---although the physics is very different, see the review \cite{Klose:2010ki}. 
The $AdS_3/CFT_2$ integrability \cite{Babichenko:2009dk,Borsato:2014hja}, see~\cite{Sfondrini:2014via,Borsato:2016hud} for reviews, also provides a fertile realisation of these exotic group-theory structures. {The studied examples are the maximally symmetric backgrounds $AdS_3\times S^3\times S^3\times S^1$ and $AdS_3\times S^3\times T^4$ with superisometry algebras $\alg{d}(2,1;\alpha)^2$ and $\alg{psu}(1,1|2)^2$, and the bonus symmetry was found in \cite{Pittelli:2014ria}, cf. \cite{Regelskis:2015xxa}}. 
Another recently integrable background is $AdS_2 \times S^2 \times T^6$, with superisometry $\alg{psu}(1,1|2)$. The holographic dual might either be a superconformal quantum mechanics, or a chiral CFT \cite{Sorokin:2011rr,Murugan:2012mf}. In \cite{Hoare:2014kma} an exact $S$-matrix theory was built, realising a centrally-extended $\alg{psu}(1|1)$ Lie superalgebra. Yangian, bonus symmetry and Bethe ansatz have been studied in \cite{Hoare:2014kma,Hoare:2015kla,Fontanella:2017rvu}.} 
{Let us mention that a boost symmetry may be found also in the context of $AdS_3/CFT_2$~\cite{Borsato:2017icj}, which seems to indicate that the boost is a rather generic feature of the $AdS/CFT$ scattering problems. }

{
 We hope that future works in the investigation of the boost symmetry will help to unveil the algebraic structure of the  underlying quantum groups.}

\section{Acknowledgments}

{We would like to thank Marius de Leeuw, Andrea Prinsloo, Alessandro Sfondrini and Bogdan Stefa\'nski for very useful insights and for encouragement. We thank Niklas Beisert, Reimar Hecht, Ben Hoare and Kostya Zarembo for invaluable discussions and several clarifications, and for detailed feedback on the manuscript. A.T. thanks Luis Miramontes for a discussion of \cite{Appadu:2017xku}.}
The work of R.B. was supported by the ERC advanced grant No 341222. A.T. thanks the STFC under the Consolidated Grant project nr. ST/L000490/1. We also thank the Galileo Galilei Institute for Theoretical Physics (GGI) for the hospitality and INFN for partial support during the completion of this work, and within the program {\it New Developments in $AdS_3/CFT_2$ Holography}.

\subsection*{Data Management}

No data beyond those presented in this paper are needed to validate its findings.


\appendix

\section{The $\psu(2|2)_{c.e.}$ symmetry}\label{sec:rev-ads5}
Here we review some known facts and we collect our conventions. Let us consider one copy of $\psu(2|2)_{c.e.}$, that contains three central elements---the hamiltonian $\gen{H}$ and two momentum-dependent charges $\gen{C},\overline{\gen{C}}$ conjugate to each other.
The non-vanishing (anti)commutation relations of this superalgebra are
\be\label{eq:comm-rel-psu22ce}
\begin{aligned}
&\{\gen{Q}_\alpha^{\ a},\overline{\gen{Q}}_b^{\ \beta}\}= \delta^a_b \gen{R}_\alpha^{\ \beta}+\delta_\alpha^\beta\gen{L}_b^{\ a}+\tfrac{1}{2}\delta^a_b\delta_\alpha^\beta \gen{H},\\
&\{\gen{Q}_\alpha^{\ a},\gen{Q}_\beta^{\ b}\}=\epsilon_{\alpha\beta}\epsilon^{ab}\gen{C},\qquad
&&\{\overline{\gen{Q}}_a^{\ \alpha},\overline{\gen{Q}}_b^{\ \beta}\}=\epsilon^{\alpha\beta}\epsilon_{ab}\overline{\gen{C}},\\
&[\gen{L}_a^{\ b},\gen{L}_c^{\ d}]=\delta^b_c\gen{L}_a^{\ d}-\delta^d_a\gen{L}_c^{\ b},\qquad
&&[\gen{R}_\alpha^{\ \beta},\gen{R}_\gamma^{\ \delta}]=\delta^\beta_\gamma\gen{R}_\alpha^{\ \delta}-\delta^\delta_\alpha\gen{R}_\gamma^{\ \beta},\\
&[\gen{L}_a^{\ b},\gen{q}_c]=\delta^b_c\gen{q}_a-\tfrac{1}{2}\delta_a^b\gen{q}_c,\qquad
&&[\gen{R}_\alpha^{\ \beta},\gen{q}_\gamma]=\delta^\beta_\gamma\gen{q}_\alpha-\tfrac{1}{2}\delta_\alpha^\beta\gen{q}_\gamma,\\
&[\gen{L}_a^{\ b},\gen{q}^c]=-\delta_a^c\gen{q}^b+\tfrac{1}{2}\delta_a^b\gen{q}^c,\qquad
&&[\gen{R}_\alpha^{\ \beta},\gen{q}^\gamma]=-\delta_\alpha^\gamma\gen{q}_\beta+\tfrac{1}{2}\delta_\alpha^\beta\gen{q}^\gamma,\\
\end{aligned}
\ee
where we used $\gen{q}$ to denote a generic element.

The above superalgebra admits an outer $\alg{sl}_2$ automorphism generated by $\gen{B},\, \gen{B}_\pm$, where $[\gen{B},\gen{B}_\pm]=\pm\gen{B}_\pm ,\ [\gen{B}_+,\gen{B}_-]=2\gen{B}$. The action of the outer $\alg{sl}_2$ on the superalgebra is 
\be\label{eq:outer-sl2}
\begin{aligned}
&[\gen{B},\gen{C}]=+\gen{C},\qquad
&&[\gen{B}_-,\gen{C}]=\gen{H},\qquad
&&&[\gen{B}_+,\gen{H}]=2\gen{C},
\\
&[\gen{B},\overline{\gen{C}}]=-\overline{\gen{C}},\qquad
&&[\gen{B}_+,\overline{\gen{C}}]=\gen{H},\qquad
&&&[\gen{B}_-,\gen{H}]=2\overline{\gen{C}},
\\
&[\gen{B},\gen{Q}_\alpha^{\ a}]=+\tfrac{1}{2}\gen{Q}_\alpha^{\ a},
&&[\gen{B}_-,\gen{Q}_\alpha^{\ a}]=\epsilon_{\alpha\beta}\epsilon^{ab}\overline{\gen{Q}}_b^{\ \beta},\qquad
\\
&[\gen{B},\overline{\gen{Q}}_a^{\ \alpha}]=-\tfrac{1}{2}\overline{\gen{Q}}_a^{\ \alpha},\qquad
&&[\gen{B}_+,\overline{\gen{Q}}_a^{\ \alpha}]=\epsilon_{ab}\epsilon^{\alpha\beta}\gen{Q}_\beta^{\ b}.
\end{aligned}
\ee
When working with the fundamental representation, the realisation for the supercharges, $\su(2)$ generators and Hamiltonian is the same as in section~\ref{sec:boost}. In the $\psu(2|2)_{c.e.}$ formulation there are also central charges, which in the fundamental  representation are  proportional to the identity matrix $\gen{C}=c\, \mathbf{1},\ \gen{C}=\bar c\, \mathbf{1}$. In our conventions we parameterise
\be
c=\bar c=\frac{ig}{2}\frac{x^+-x^-}{\sqrt{x^-x^+}}=-g\, \sin\frac{p}{2}.
\ee
One may equip $\psu(2|2)_{c.e.}$ with a coproduct to obtain the structure of a bialgebra. We prefer to use the coproduct in the {}{most symmetric frame}~\cite{Arutyunov:2009ga}
\be\label{eq:Delta-psu(2|2)ce}
\begin{aligned}
&\Delta{\gen{H}}=\gen{H}\otimes 1+1\otimes\gen{H},&& \\
&\Delta{\gen{L}_a^{\ b}}=\gen{L}_a^{\ b}\otimes 1+1\otimes\gen{L}_a^{\ b},\qquad
&& \Delta{\gen{R}_\a^{\ \b}}=\gen{R}_\a^{\ \b}\otimes 1+1\otimes\gen{R}_\a^{\ \b},\\
&\Delta{\gen{C}}=\gen{C}\otimes e^{-\frac{i}{2}p}+e^{\frac{i}{2}p}\otimes\gen{C},\qquad
&&\Delta{\overline{\gen{C}}}=\overline{\gen{C}}\otimes e^{\frac{i}{2}p}+e^{-\frac{i}{2}p}\otimes\overline{\gen{C}},\\
&\Delta{\gen{Q}_\a^{\ a}}=\gen{Q}_\a^{\ a}\otimes e^{-\frac{i}{4}p}+e^{\frac{i}{4}p}\otimes\gen{Q}_\a^{\ a},\qquad
&&\Delta{\overline{\gen{Q}}_a^{\ \a}}=\overline{\gen{Q}}_a^{\ \a}\otimes e^{\frac{i}{4}p}+e^{-\frac{i}{4}p}\otimes\overline{\gen{Q}}_a^{\ \a}.
\end{aligned}
\ee
Because of the braiding factors $e^{ip}$, the coproducts for the supercharges and for the central charges $\gen{C},\overline{\gen{C}}$ are not trivial. This makes sure e.g. that the eigenvalue of $\Delta(\gen{C})$ is $-g\sin\tfrac{p_1+p_2}{2}$, i.e. the {}{momenta} are added. The above coproducts may be summarised as
\be
\Delta{\gen{q}^{A}} = \gen{q}^A\otimes e^{-\frac{i}{2}[A]p}+e^{\frac{i}{2}[A]p}\otimes \gen{q}^A,
\ee
where $\gen{q}^A$ stands for any of the generators, and $[A]$ is the corresponding eigenvalue under the Cartan of the outer $\alg{sl}_2$.

In the fundamental representation it is also possible to derive an $R$-matrix by demanding the compatibility with the symmetry algebra
\be
\Delta^{op}(\gen{q}^A)R=R\Delta(\gen{q}^A),\qquad \forall A.
\ee
Here $^{op}$ denotes the opposite coproduct and we take, $\Delta^{op}(\gen{q})=\Pi_g\cdot\Delta(\gen{q})\cdot\Pi_g$ and $\Pi_g$ is the graded permutation.\footnote{On the matrix realisation we explicitly swap $p_1\leftrightarrow p_2$ after applying the graded permutation, so that the momenta $p_1,p_2$ always label the first and the second space respectively.} Compatibility with all elements of $\psu(2|2)_{c.e.}$ fixes the $R$-matrix uniquely up to an overall scalar factor in the fundamental representation~\cite{Beisert:2005tm}. This is an explicit $16\times 16$ matrix, and in our conventions it reads as
\be\label{eq:S-mat}
R=\sum_{k=1}^{10}a_k\Lambda_k,
\ee
where $\Lambda_k$ are given after Equation (3.84) of~\cite{Arutyunov:2009ga}, while the momenta-dependent coefficients $a_k$ are
\be
\begin{aligned}
& a_1=1,\\
& a_2=2\frac{ x^-_1}{x^+_1}\ \frac{  x^-_2x^+_1-1}{x^-_1x^-_2-1}\ \frac{ x^+_1-x^+_2}{x^-_1-x^+_2}-1,\\
& a_3=\frac{x^-_2-x^+_1}{x^-_1-x^+_2}\left(\frac{x^+_2}{x^-_2}\right)^{\frac{1}{2}} \left( \frac{x^+_1}{x^-_1} \right)^{-\frac{1}{2}},\\
& a_4=\frac{ \left(2 x^-_1x^-_2 \left(x^+_2\right)^2-\left(x^-_1x^-_2+1\right) \left(x^-_2+x^+_1\right)x^+_2+2 x^-_2 x^+_1\right)}{\left(x^-_1x^-_2-1\right) \left(x^-_1-x^+_2\right)x^+_2}
\left(\frac{x^+_2}{x^-_2}\right)^{\frac{1}{2}} \left( \frac{x^+_1}{x^-_1} \right)^{-\frac{1}{2}},\\
& a_5=\frac{x^+_1-x^+_2}{x^-_1-x^+_2} \left(\frac{x^+_1}{x^-_1} \right)^{-\frac{1}{2}},\\
& a_6=\frac{x^-_2-x^-_1}{x^+_2-x^-_1}\left(\frac{x^+_2}{x^-_2}\right)^{\frac{1}{2}},\\
& a_7=-a_8=\frac{i \gamma_1\gamma_2}{x^-_1x^-_2-1}\ \frac{x^+_1-x^+_2}{x^-_1-x^+_2}\ \left(\frac{x^+_1}{x^-_1}\right)^{-\frac{3}{4}}\left(\frac{x^+_2}{x^-_2}\right)^{-\frac{1}{4}},\\
& a_9=a_{10}=-\frac{i\gamma_1 \gamma_2}{x^-_1-x^+_2}\ \left(\frac{x^+_1}{x^-_1}\right)^{-\frac{1}{4}}\left(\frac{x^+_2}{x^-_2}\right)^{\frac{1}{4}}.
\end{aligned}
\ee
The normalisation chosen here is arbitrary. 

\vspace{12pt}

In~\cite{Beisert:2007ds} a Yangian for $\psu(2|2)_{c.e.}$ was constructed. The level 0 of the Yangian coincides with the superalgebra presented above; at level-1 one has generators $\hat{\gen{q}}^A$ so that commutation relations are
\be\label{eq:commrel-yang}
[\gen{q}^A,\gen{q}^B]=f^{AB}_{C}\ \gen{q}^C,\qquad
[\gen{q}^A,\hat{\gen{q}}^B]=f^{AB}_{C}\ \hat{\gen{q}}^C.
\ee
Additional Serre relations, that we omit, give the Yangian in Drinfeld's first realisation, see~\cite{Beisert:2007ds}.
By multiplication with the spectral parameter $\hat u$ we obtain the level 1 in the evaluation representation
\be
\hat{\gen{q}}^A\sim \hat u \gen{q}^A,\qquad\qquad
\hat u=\frac{\hbar}{2}\left( x^++\frac{1}{x^+}+x^-+\frac{1}{x^-} \right).
\ee
The Yangian contains an additional deformation parameter $\hbar$ which can be set to 1, as we will do in the remaining sections.

In order to write down coproducts for the charges at level 1, one considers the algebra extended by the outer $\alg{sl}_2$ automorphism, for which it is possible to write a non-degenerate invariant bilinear form\footnote{The Killing form of $\psu(2|2)_{c.e.}$ vanishes, and it is not possible to find a non-degenerate bilinear form \cite{Spill:2007bia} (see also \cite{Dobrev:2009bb}). The Killing form of the extended algebra is also degenerate.}
\be
G^{AB}= g^{AB}+s^{AB},\qquad\qquad G^{AB}\equiv G(T^A,T^B).
\ee
Here $g^{AB}$ are the components of a non-degenerate bilinear form on the extended algebra
\be
\begin{aligned}
&g(\gen{Q}_\alpha^{\ a},\overline{\gen{Q}}_b^{\ \beta})=\delta_\alpha^\beta\, \delta^a_b,
&&g(\gen{L},\gen{L})=-\tfrac{1}{2},
&&&g(\gen{L}_+,\gen{L}_-)=-1,
&&&&g(\gen{C},\gen{B}_-)=-1,\\
&g(\gen{H},\gen{B})=1,
&&g(\gen{R},\gen{R})=+\tfrac{1}{2},
&&&g(\gen{R}_+,\gen{R}_-)=+1,
&&&&g(\overline{\gen{C}},\gen{B}_+)=+1,
\end{aligned}
\ee
where $\gen{L}\equiv \gen{L}_1^{\ 1}$, $\gen{L}_+\equiv\gen{L}_1^{\ 2}$, $\gen{L}_-\equiv\gen{L}_2^{\ 1}$ and similarly for $\gen{R}_\alpha^{\ \beta}$.
The bilinear form $s^{AB}$ is non-vanishing only on $\alg{sl}_2$, and we set it to 0.
Then the coproducts for the level-1 generators  are obtained by
\be\label{eq:Delta-qhat}
\Delta(\hat{\gen{q}}^A) = \hat{\gen{q}}^A \otimes e^{-\frac{i}{2}[A]p}+ e^{\frac{i}{2}[A]p} \otimes \hat{\gen{q}}^A +\frac{i\hbar}{g} f^A_{BC}\ \ e^{\frac{i}{2}[C]p} \gen{q}^B \otimes e^{-\frac{i}{2}[B]p}\gen{q}^C.
\ee
We have lowered one index of the structure constants with the inverse of the bilinear form $f^A_{BC}\equiv f^{DA}_CG_{BD}$.
One can check that this coproduct is an algebra homomorphism for~\eqref{eq:commrel-yang} and that the level-1 generators are symmetries of the $R$-matrix.

\vspace{12pt}

In~\cite{Matsumoto:2007rh,Beisert:2007ty} an additional {}{secret symmetry} of the $R$-matrix was discovered. In the fundamental representation on one-particle states it is realised as
\be\label{eq:secret}
\hat{\gen{B}}=\frac{\hbar}{4}\left(x^++x^--\frac{1}{x^+}-\frac{1}{x^-}\right)\ \mathbf{1}_g,
\ee
where $\mathbf{1}_g=\text{diag}(1,1,-1,-1)$ is the graded identity.
The secret symmetry commutes with all bosonic charges of $\psu(2|2)_{c.e.}$, and commutation relations with supercharges are compatible with
\be
\begin{aligned}\label{eq:comm-B-Q}
[\hat{\gen{B}},\gen{Q}_\alpha^{\ a}]&=-\hat{\gen{Q}}_\alpha^{\ a}-2\hbar\, \cos\frac{p}{2}\epsilon_{\alpha\beta}\epsilon^{ab}\overline{\gen{Q}}{}_b^{\ \beta},\\
[\hat{\gen{B}},\overline{\gen{Q}}_a^{\ \alpha}]&=+\hat{\overline{\gen{Q}}}{}_a^{\ \alpha}+2\hbar\, \cos\frac{p}{2}\epsilon^{\alpha\beta}\epsilon_{ab}\gen{Q}_\beta^{\ b}.
\end{aligned}
\ee
A coproduct compatible with these commutation relations is
\be\label{eq:DeltaB}
\Delta(\hat{\gen{B}})=\hat{\gen{B}}\otimes 1+1\otimes \hat{\gen{B}}
+\frac{i\hbar}{g}\left(e^{-\frac{i}{4}p}\, \gen{Q}_\a^{\ a}\otimes e^{-\frac{i}{4}p}\, \overline{\gen{Q}}_a^{\ \a}
+e^{\frac{i}{4}p}\, \overline{\gen{Q}}_a^{\ \a}\otimes e^{\frac{i}{4}p}\, \gen{Q}_\a^{\ a}\right),
\ee
and a direct computation shows that it is indeed a symmetry of the $R$-matrix.

\section{Solving the $q$-Serre relations}\label{app:q-aff}
The contraction of $U_q(\widehat{\alg{sl}_2})$ must be taken carefully. In fact, if one applies na\"ively the contraction to the $q$-Serre relations, one may be tricked into writing down a set of commutation relations which are wrong, since they do not satisfy the Jacobi identity. 

Here we take a safer path. First we {}{solve} the $q$-Serre relations for $U_q(\widehat{\alg{sl}_2})$, so that we have all commutators at our disposal; the contraction can then be safely implemented.
Let us define
\be
F^+(m,n)\equiv [\gen{e}^+_m,\gen{e}^+_n].
\ee
We have
\be
F^+(m,n)=[\gen{e}^+_m,\gen{e}^+_n]_{q^2}-(1-q^{2})\gen{e}^+_n\gen{e}^+_m=-[\gen{e}^+_{n+1},\gen{e}^+_{m-1}]_{q^2}-(1-q^{2})\gen{e}^+_n\gen{e}^+_m,
\ee
where in the last step we used the last equation in~\eqref{eq:def-rel-Uqsl2}. Moreover
\be
F^+(m-1,n+1)=-q^{-2}[\gen{e}^+_{n+1},\gen{e}^+_{m-1}]_{q^2}-(1-q^{-2})\gen{e}^+_{n+1}\gen{e}^+_{m-1}.
\ee
Putting the two together we find
\be\label{eq:rec-relFp-1}
F^+(m-1,n+1)=q^{-2}F^+(m,n)-(1-q^{-2})(\gen{e}^+_{n}\gen{e}^+_{m}+\gen{e}^+_{n+1}\gen{e}^+_{m-1}).
\ee
Notice that this relation connects different points of the lattice $(m,n)$ which belong to the same line $m+n=$costant.

Let us now consider the case of $n+m$ even. We call $n+m=2s, n-m=2d$, so that $n=s+d, m=s-d$. The case $d=0$ is trivial $F^+(s,s)=0$. We set $d>0$ for definiteness. Using the above relation it is easy to show by induction that 
\be
F^+(s-d,s+d)=-(1-q^{-2})\left( q^{-2(d-1)} \gen{e}^+_{s}\gen{e}^+_{s}+\gen{e}^+_{s+d}\gen{e}^+_{s-d}\right)
-(1-q^{-4})\sum_{\ell =1}^{d-1}q^{-2(d-1-\ell)}\gen{e}^+_{s+\ell}\gen{e}^+_{s-\ell}.
\ee
Similarly, for the case of $n+m$ odd. We call $n+m=2s+1, n-m=2d+1$, so that $n=s+d+1, m=s-d$. Directly from~\eqref{eq:def-rel-Uqsl2} we derive 
\be
[\gen{e}^+_{s},\gen{e}^+_{s+1}]=F^+(s,s+1)=-(1-q^{-2})\gen{e}^+_{s+1}\gen{e}^+_s,
\ee
and then we use it to prove for $d>0$
\be
F^+(s-d-1,s+d+2)= -(1-q^{-2})\gen{e}^+_{s+d+2}\gen{e}^+_{s-d-1}-(1-q^{-4})\sum_{\ell=0}^dq^{-2(d-\ell)}\gen{e}^+_{s+1+\ell}\gen{e}^+_{s-\ell}.
\ee
The same formulas are valid for $\gen{e}^-_m$ upon sending $q\to q^{-1}$. Let us notice that we could change a bit the derivation such that the operators $\gen{e}^+_j$ appear in different orders in the products. In fact from the two equations
\be
\begin{aligned}
F^+(m,n)&=-[\gen{e}^+_n,\gen{e}^+_m]_{q^2}+(1-q^2)\gen{e}^+_m\gen{e}^+_n=-[\gen{e}^+_{m+1},\gen{e}^+_{n-1}]_{q^2}+(1-q^2)\gen{e}^+_m\gen{e}^+_n,\\
F^+(m+1,n-1)&=q^{-2}[\gen{e}^+_{m+1},\gen{e}^+_{n-1}]_{q^2}+(1-q^{-2})\gen{e}^+_{m+1}\gen{e}^+_{n-1},
\end{aligned}
\ee
we obtain a new relation
\be
F^+(m-1,n+1)=q^2F^+(m,n)+(1-q^2)(\gen{e}^+_{m-1}\gen{e}^+_{n+1}+\gen{e}^+_{m}\gen{e}^+_{n}).
\ee
The above relation may be obtained from~\eqref{eq:rec-relFp-1} after sending $q\to q^{-1}$, swapping the order of all products $\gen{e}^+_i\gen{e}^+_j\to \gen{e}^+_j\gen{e}^+_i$, and changing the sign of the last term.
One can repeat the previous derivation, so that the result for $[\gen e_m^+,\gen e_n^+]$ in each case may be written in two equivalent ways
\be
\begin{aligned}
&m+n=\text{even},\ m<n:\ &&-(1-q^{-2})\left( q^{2+m-n} \gen{e}^+_{\frac{m+n}{2}}\gen{e}^+_{\frac{m+n}{2}}+\gen{e}^+_{n}\gen{e}^+_{m}\right)\\
& && \quad-(1-q^{-4})\sum_{\ell =1}^{\frac{n-m}{2}-1}q^{2+m-n+2\ell}\gen{e}^+_{\frac{m+n}{2}+\ell}\gen{e}^+_{\frac{m+n}{2}-\ell}\\
& && =(1-q^{2})\left( q^{-2-m+n} \gen{e}^+_{\frac{m+n}{2}}\gen{e}^+_{\frac{m+n}{2}}+\gen{e}^+_{m}\gen{e}^+_{n}\right)\\
& && \quad +(1-q^{4})\sum_{\ell =1}^{\frac{n-m}{2}-1}q^{-2-m+n-2\ell}\gen{e}^+_{\frac{m+n}{2}-\ell}\gen{e}^+_{\frac{m+n}{2}+\ell},
\\
&n=m+1:\ &&-(1-q^{-2})\gen{e}^+_{m+1}\gen{e}^+_m\\
& &&=(1-q^{2})\gen{e}^+_m\gen{e}^+_{m+1},
\\
&m+n=\text{odd},\ m<n+2:\ &&-(1-q^{-2})\gen{e}^+_{n}\gen{e}^+_{m}
\\
& &&\quad-(1-q^{-4})\sum_{\ell=0}^{\frac{n-m-1}{2}-1}q^{2\ell-n+m+3}\gen{e}^+_{\frac{m+n-1}{2}+1+\ell}\gen{e}^+_{\frac{m+n-1}{2}-\ell}\\
& &&= (1-q^{2})\gen{e}^+_{m}\gen{e}^+_{n}\\
& &&\quad+(1-q^{4})\sum_{\ell=0}^{\frac{n-m-1}{2}-1}q^{-2\ell+n-m-3}\gen{e}^+_{\frac{m+n-1}{2}-\ell}\gen{e}^+_{\frac{m+n-1}{2}+1+\ell}.
\end{aligned}
\ee
We can therefore freely choose between the two sets of equations.
Using the first or the second set of formulas to compute commutators of $\gen H_m,\gen J_m,\gen P_m$ we obtain
\be
\begin{aligned}
[\gen H_m,\gen J_n]&=-\frac{1}{2}\left(\bpsi^+_{m+n}-\bpsi^-_{m+n}\right)-\frac{1}{2} \ \text{sign}(n-m)\sum_{\ell=-\frac{|n-m|}{2}}^{\frac{|n-m|}{2}-1}\gen H_{\frac{n+m}{2}+\ell}\gen H_{\frac{n+m}{2}-\ell}\\
&=-\frac{1}{2}\left(\bpsi^+_{m+n}-\bpsi^-_{m+n}\right)-\frac{1}{2}\ \text{sign}(n-m)\left(\sum_{\ell=-\frac{n-m}{2}}^{\frac{n-m}{2}}\gen H_{\frac{n+m}{2}+\ell}\gen H_{\frac{n+m}{2}-\ell} -\gen H_n\gen H_m\right),
\end{aligned}\ee
where we do not need to specify whether $m+n$ is even or odd: in the former case the index $\ell\in\mathbb{Z}$, in the latter $\ell\in\frac{1}{2}+\mathbb{Z}$, and in both cases we take an incremental step of 1. Notice that when $m>n$ we take the sum to run in the opposite direction, i.e. from positive to negative.
The commutator of two $\gen J$'s is a bit more involved. We first implement the contraction on all above expressions, and then take the semisum of the two equivalent results in each case. We obtain
\be
\begin{aligned}
[\gen J_m,\gen J_n]=& \frac{1}{2}\ \text{sign}(m-n)\Bigg(\sum_{\ell=-\frac{n-m}{2}}^{\frac{n-m}{2}}\left(\gen H_{\frac{m+n}{2}+\ell}\gen J_{\frac{m+n}{2}-\ell}+\gen J_{\frac{m+n}{2}+\ell}\gen H_{\frac{m+n}{2}-\ell}\right)\\
&\qquad -\frac{1}{2}\left(\gen H_m\gen J_n+\gen J_m\gen H_n+\gen H_n\gen J_m+\gen J_n\gen H_m\right)\Bigg).
\end{aligned}
\ee
These results can be rewritten as in~\eqref{eq:comm-rel-q-aff-poin}.
As a double-check, we have written a {}{Mathematica} code that computes Jacobi for the above commutation relations at fixed (generic) levels.

\bibliographystyle{nb}
\bibliography{biblio}{}

\end{document}